\documentclass[12pt]{article}
\usepackage[utf8]{inputenc}
\usepackage[usenames,dvipsnames]{color}
\usepackage{amsmath}
\usepackage{amssymb}
\usepackage{mathptmx}
\usepackage{authblk}

\usepackage[a4paper, margin=1in]{geometry}

\usepackage[left]{lineno}

\usepackage{multirow}
\usepackage{array}
\usepackage{rotating}
\usepackage{booktabs}%

\usepackage[breaklinks,colorlinks,citecolor=RoyalBlue,linkcolor=magenta]{hyperref}

\usepackage[super,compress,comma,sectionbib]{natbib}
\usepackage{graphicx}
\usepackage{subcaption}
\usepackage[labelfont=bf]{caption}
\DeclareCaptionLabelSeparator{bar}{ \textbar{} }
\newcommand{\arcsec}{$^{\prime\prime}$}
\newcommand{\arcmin}{$^{\prime}$}
\newcommand{\arcdeg}{$^{\circ}$}
\newcommand{\micron}{$\mu$m}
\newcommand\farcs{\mbox{$.\!^{\prime\prime}$}}

\def\jnl@style{\it}
\def\aaref@jnl#1{{\jnl@style#1}}

\newcommand{\araa}{Annu. Rev. Astron. Astrophys.}   
\newcommand{\aj}{Astron. J.}   
\newcommand{\apj}{Astrophys. J.}   
\newcommand{\apjl}{Astrophys. J. Lett.}   
\newcommand{\apjs}{Astrophys. J. Suppl. Ser.}   
\newcommand{\aap}{Astron. Astrophys.}   
\newcommand{\aapr}{Astron. Astrophys. Rev.}   
\newcommand{\mnras}{Mon. Not. R. Astron. Soc.}   
\newcommand{\nat}{Nature} 
\newcommand{\pasa}{Publ. Astron. Soc. Aust.}   
\newcommand{\pasp}{Publ. Astron. Soc. Pac.}   

\renewcommand{\figurename}{Fig.}
\newcommand{\mthc}{\mbox{CH$_3$CN}}

\newcommand{\tmthc}{\mbox{$^{13}$CH$_3$CN}}
\newcommand{\mf}{\mbox{CH$_3$OCHO}}

\newcommand{\kms}{\mbox{km\,s$^{-1}$}}
\newcommand{\cc}{\mbox{cm$^{-3}$}}
\newcommand{\lsun}{\mbox{$L_\odot$}}
\newcommand{\msun}{\mbox{$M_\odot$}}

\title{\textbf{Observations of high-order multiplicity in a high-mass stellar protocluster}}

\author[1*]{Shanghuo Li}
\affil[1]{\small Max Planck Institute for Astronomy, Heidelberg, Germany; \url{shanghuo.li@gmail.com}}

\author[2,3]{Patricio Sanhueza}
\affil[2]{National Astronomical Observatory of Japan, National Institutes of Natural Sciences, Tokyo, Japan}
\affil[3]{Department of Astronomical Science, School of Physical Science, SOKENDAI (The Graduate University for Advanced Studies), Tokyo, Japan}

\author[1]{Henrik Beuther}

\author[4]{Huei-Ru Vivien Chen}
\affil[4]{Institute of Astronomy and Department of Physics, National Tsing Hua University, Hsinchu, Taiwan}

\author[5]{Rolf Kuiper}
\affil[5]{Faculty of Physics, University of Duisburg-Essen, Duisburg, Germany}

\author[4]{Fernando A. Olguin}

\author[6]{Ralph E. Pudritz}
\affil[6]{Origins Institute and Department of Physics and Astronomy, McMaster University, Hamilton, Ontario, Canada}

\author[7]{Ian W. Stephens}
\affil[7]{Department of Earth, Environment, and Physics, Worcester State University, Worcester, MA, USA}

\author[8]{Qizhou Zhang}
\affil[8]{Center for Astrophysics, Harvard \& Smithsonian, Cambridge, MA, USA}

\author[2,3]{Fumitaka Nakamura}

\author[9]{Xing Lu}
\affil[9]{Shanghai Astronomical Observatory, Chinese Academy of Sciences, Shanghai, People’s Republic of China}

\author[10]{Rajika L. Kuruwita}
\affil[10]{Heidelberg Institute for Theoretical Studies, Heidelberg, Germany}

\author[11]{Takeshi Sakai}
\affil[11]{Graduate School of Informatics and Engineering, The University of Electro-Communications, Tokyo, Japan}

\author[1]{Thomas Henning}

\author[2]{Kotomi Taniguchi}

\author[12]{Fei Li}
\affil[12]{School of Astronomy and Space Science, Nanjing University, People's Republic of China}

\date{}
 
\begin{document}

\maketitle

{\color{blue}
\noindent
The dominant mechanism forming multiple stellar systems in the high-mass regime (M$_\ast \gtrsim $ 8 $M_{\odot}$) remained unknown  because direct imaging of multiple protostellar systems at early phases of high-mass star formation is very challenging. High-mass stars are expected to form in clustered environments containing binaries and higher-order multiplicity systems. So far only a few high-mass protobinary systems, and no definitive higher-order multiples, have been detected. 
Here we report the discovery of one quintuple, one quadruple, one triple and four binary protostellar systems simultaneously forming in a single high-mass protocluster, G333.23--0.06, using Atacama Large Millimeter/submillimeter Array high-resolution observations.  
We present a new example of a group of gravitationally bound  binary and higher-order multiples during their early formation phases in a protocluster.  
This provides the clearest direct measurement of the initial configuration of  primordial high-order multiple systems, with implications for the in situ multiplicity and its origin. 
We find that the binary and higher-order multiple systems, and their parent cores,  show no obvious sign of disk-like kinematic structure. 
We conclude that the observed fragmentation into binary and higher-order multiple systems can be explained by core fragmentation, indicating its crucial role in establishing the multiplicity during high-mass star cluster formation. 
}

\begin{figure}[!h]
\centering
\includegraphics[width=1\textwidth]{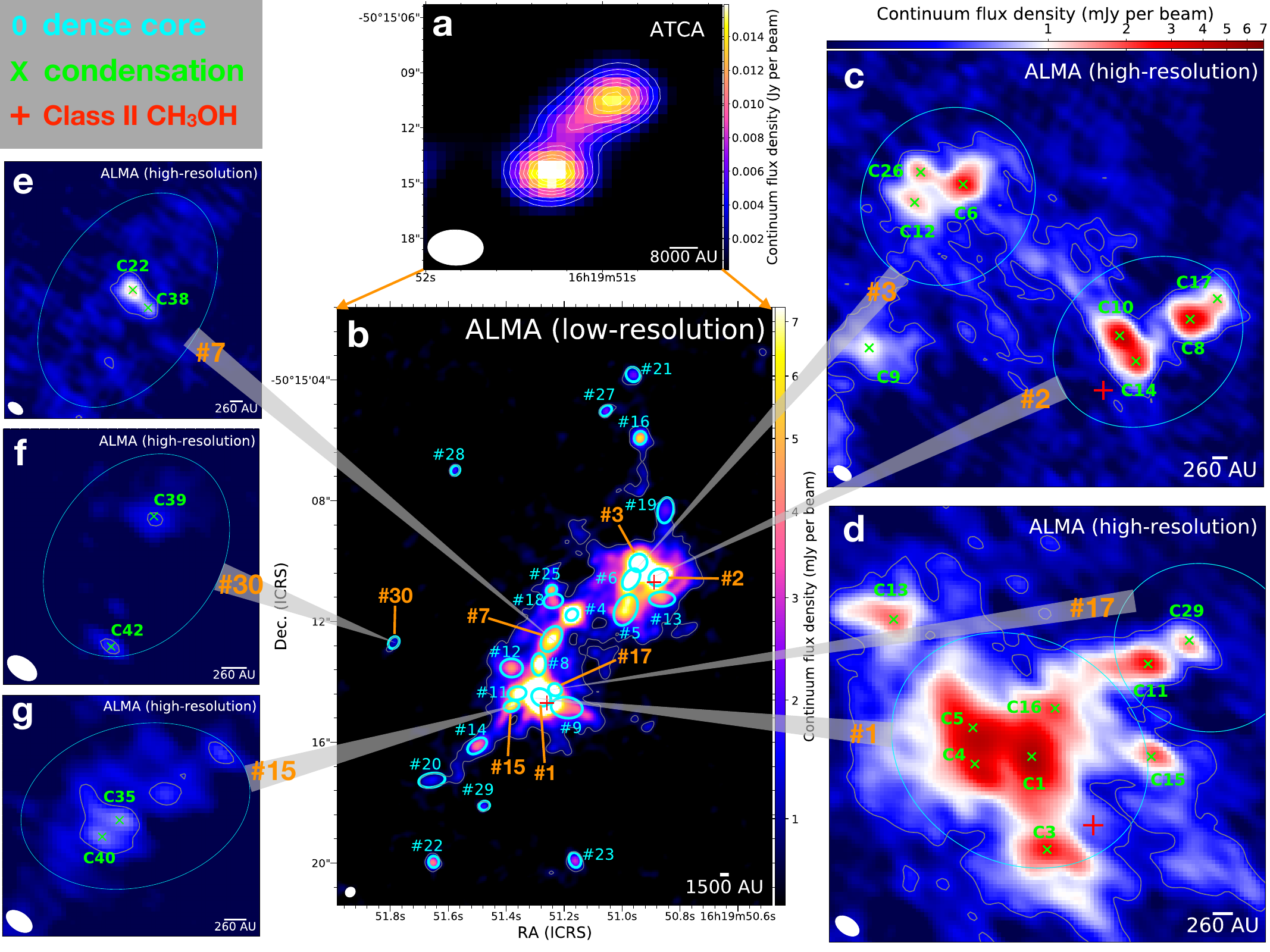}
\caption{
\textbf{Continuum images of ATCA (3.3~mm), ALMA low-resolution (1.3~mm), and ALMA  high-resolution (1.3~mm) observations.} \textbf{a}, ATCA 3.3~mm continuum image ($\theta \sim$ 2.42\arcsec{}). The contour levels are [5, 8, 11, 14,17, 20, 23, 26, 29] $\times$ $\sigma$, where root mean squared (rms) noise of image is $\sigma$=0.6~mJy beam$^{-1}$. \textbf{b}, ALMA low-resolution ($\theta \sim$ 0.32\arcsec{}) 1.3~mm continuum image. The contour is the 7 $\sigma$, where $\sigma$=0.16~mJy beam$^{-1}$. The ellipses show the identified cores based on the low-resolution  continuum image. The red pluses are the Class II  CH$_{3}$OH maser \citep{Caswell2011}.  
\textbf{c--g}, ALMA high-resolution ($\theta \sim$ 0.05\arcsec{}) 1.3~mm continuum image for multiple systems. 
The green crosses present the identified condensations based on the ALMA high-resolution continuum image. 
The contour is the 7$\sigma$, where $\sigma$=0.05~mJy beam$^{-1}$. 
The dense cores (\#1, \#2, \#3, ...) and condensations (C1, C2, C3, ...) are numbered in order of descending integrated intensity. 
\textbf{c}, Dense core \#2 fragments into quadruple condensation system (C8, C10, C14, C17), and  dense core \#3 fragments into triple condensation system (C6, C12, C26). 
\textbf{d}, Dense core \#1 fragments into quintuple condensation system (C1, C3, C4, C5, C16), and dense core \#17 fragments into binary condensation (C11-C29). 
\textbf{e}, Dense core \#7 fragments into binary condensation (C22-C38). 
\textbf{f}, Dense core \#30 fragments into binary condensation (C39-C42). 
\textbf{g}, Dense core \#15 fragments into binary condensation (C35-C40). 
The white ellipses in the lower left corner of each panel denote the synthesized beam of continuum images. 
ICRS, International Celestial Reference System. 
}
\label{fig:fig1}
\end{figure}


High-mass stars in the Milky Way are overwhelmingly ($>$80\%; refs.\citep{Kouwenhoven2007,Zinnecker2007,Mason2009,Duchene2013,Sana2014,Offner2022}) found in binaries  or higher-order multiplicity systems that play a key role in governing cluster dynamics and stellar evolution \citep{Portegies-Zwart2010,Sana2012}.  
However, it is yet unclear whether they are predominantly formed from in situ fragmentation at various scales (e.g., disks\citep{Bonnell1994,Kratter2010}, cores\citep{Larson1972,Offner2010}, or filaments\citep{Pineda2015})  or subsequent stellar capture in clusters\citep{Bate2003} because direct measurements of their initial configuration and properties at the early phases of cluster formation have been unattainable \citep{Reipurth2014,Beltran2016a,Brogan2016,Beuther2017,Orozco-Aguilera2017,Ilee2018,Zapata2019,Zhang2019,Guzman2020,Tanaka2020,Beltran2021,Cyganowski2022,Olguin2022}.

We report the direct imaging of 1 quintuple, 1 quadruple, 1 triple, and 4 binary systems in the high-mass protocluster G333.23--0.06 (hereafter G333) by using Atacama Large Millimeter/submillimeter Array (ALMA) long-baseline observations (Fig.~\ref{fig:fig1} and Extended Data Fig.~\ref{fig:highcont}). 
These binary and higher-order systems are detected in the high-resolution (angular resolution of $\theta \sim $ 0.05\arcsec{}, equivalent to 260~au at the source distance of 5.2 kpc; ref.\citep{Whitaker2017}) 1.3~mm dust continuum image. 
The detected condensations have radii between 153 and 678 au (Table~\ref{tab1}). 
We refer to both binary and higher-order systems simply as multiple systems in what follows, and we only make an explicit distinction between the two when necessary.

The projected separations of these multiple systems are between 327 and 1406 au, with a mean value of 731 au (Extended Data Fig.~\ref{fig:separation}), in good agreement with the typical projected separation of 700~au in the simulation of multiple star formation via core fragmentation \citep{Kuruwita2023}.  
The ambient gas masses ($M_{\rm amb}$) of these multiple systems range from 0.19 to 1.47 \msun\  on the basis of the thermal dust emission (\hyperref[sec:methods]{Methods}). 
These masses are regarded as lower limits because 
the observations suffered from missing flux in the interferometer data. 
We focus primarily on the multiple systems that are embedded in a single dense core (typical radius of $\sim$2100 au; Fig.~\ref{fig:fig1}). 
The quintuple system consists of a small group of condensations (C1-C4-C5-C16), which is tightly connected as seen in dust continuum emission, and a condensation (C3) slightly separated. That is nevertheless part of the original parental core (Fig.~\ref{fig:fig1}).   
The quadruple system includes two binary configurations (C10-C14 and C8-C17), and the triple system composes three slightly separated condensations (C6, C12, and C26). 
The binary systems are C11-C29, C22-C38, C39-C42, and C35-C40.

In addition to the projected proximity on the sky of observed condensations, the line-of-sight velocity is another important diagnostic tool to determine whether members of a multiple system are physically associated. 
All members of each multiple system have similar centroid velocities (Extended Data Fig.~\ref{fig:ch3oh} and \hyperref[sec:methods]{Methods}), indicating that the members are physically associated and share a common origin.

\begin{figure}[!h]
\centering
\includegraphics[width=1\textwidth]{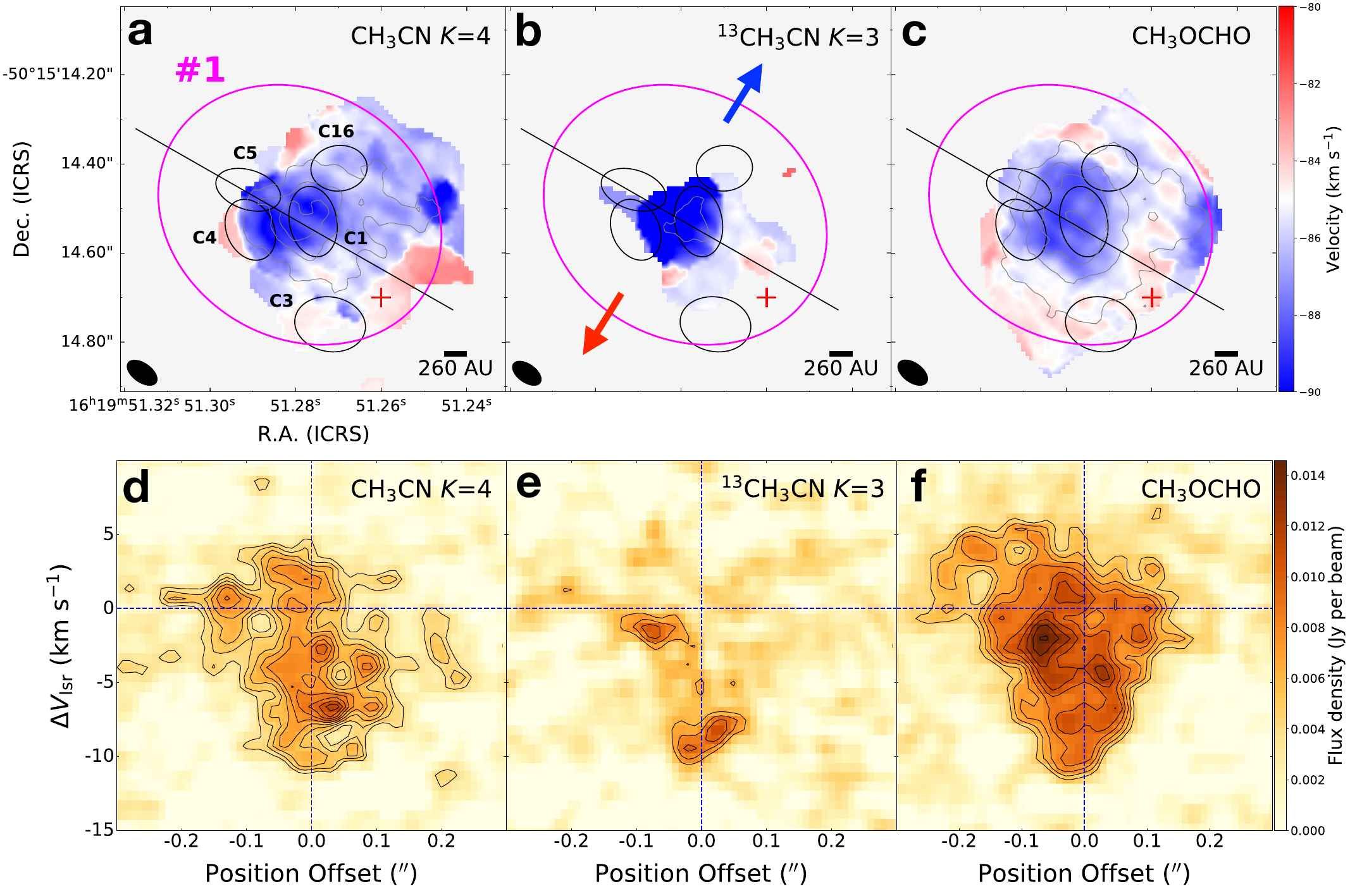}
\caption{
\textbf{Examples of intensity-weighted velocity maps and position-velocity (PV) diagrams of molecular lines for the quintuple system}. Columns left, middle, and right present the results derived from the CH$_3$CN $12_4-11_4$, $^{13}$CH$_3$CN $13_3-12_3$, and CH$_3$OCHO $20_{0,20}-19_{0,19}$, respectively.  
\textbf{a--c}, We present the intensity-weighted velocity maps derived from the ALMA high-resolution data of CH$_3$CN $12_4-11_4$ (\textbf{a}), $^{13}$CH$_3$CN $13_3-12_3$  (\textbf{b}), and CH$_3$OCHO $20_{0,20}-19_{0,19}$ (\textbf{c}). 
The black and magenta ellipses show the condensations and their parent cores, respectively. 
The red plus marks the Class II CH$_{3}$OH maser position. 
The blue and red arrows show the directions of the main outflow seen in the SiO emission from the ALMA low-resolution data. 
The grey contours show the velocity-integrated intensity maps with levels of [5, 10, 15, 20]$\times \sigma_{\rm rms}$, where $\sigma_{\rm rms}$ is 7.5~mJy~beam$^{-1}$ \kms. 
The black ellipses in the lower left corner of each panel denote the synthesized beam of continuum images.   
The remaining velocity maps are presented in Extended Data Figs.~\ref{fig:lowmom1}--\ref{fig:highmom11}. 
\textbf{d--f}, We show the PV diagram maps derived from the high-resolution data of CH$_3$CN $12_4-11_4$ (\textbf{d}), $^{13}$CH$_3$CN $13_3-12_3$ (\textbf{e}), and CH$_3$OCHO $20_{0,20}-19_{0,19}$ (\textbf{f}).  Contours levels start at 4$\sigma_{\rm rms}$ and increase in step of  1$\sigma_{\rm rms}$ interval, where $\sigma_{\rm rms}$ is 1.2~mJy~beam$^{-1}$. 
The cut lines of the PV diagram are indicated in \textbf{a--c} with black lines. 
The blue dashed lines in the vertical and horizontal directions show the position of the dense core \#1 and its systemic velocity of V$_{\rm lsr}$ = -85 \kms.
}
\label{fig:fig2}
\end{figure}

\vspace{0.5em}
\noindent{\textbf{Multiple systems formed via core fragmentation}}

The parent cores of the multiple systems are revealed in the lower-resolution ($\theta \sim$ 0.32\arcsec{}, equivalent to 1664 au) 1.3~mm dust continuum image (Fig.~\ref{fig:fig1} and Extended Data  Fig.~\ref{fig:lowcont}).  
The dense cores are identified with radii ranging from  927 to 3443 au (Table~\ref{tab1}).  
The multiple condensation systems are embedded in the dense cores. 
Each condensation likely harbors embedded  protostellar object(s) as evidenced by the presence of hot ($T_{\rm gas}$ =  108--532 K) and warm gas resulting from internal heating (\hyperref[sec:methods]{Methods}), except for C39 and C42 where no significant molecular warm transitions (that is, upper energy level of $E_{u}/k>$ 45 K) are detected.

There is no obvious sign of a disk-like kinematic structure around any of the multiple systems and their parent cores in any of the lines we examined (\hyperref[sec:methods]{Methods}), including the typical disk tracers, e.g., CH$_3$CN, $^{13}$CH$_3$CN, and CH$_3$OCHO (Fig.~\ref{fig:fig2} and Extended Data Figs.~\ref{fig:lowmom1}--\ref{fig:highmom11}), and dense gas tracer, e.g., CH$_3$OH, $^{13}$CH$_3$OH, SO$_2$, SO, HC$_3$N, HNCO, NH$_2$CHO, H$_2$CO, and H$_2^{13}$CO. 
Meanwhile, the presence of the SiO outflows indicates that the accretion disk is not viewed face-on, which rules out the scenario of a weak velocity gradient resulting from projection of a face-on geometry (Fig.~\ref{fig:fig2} and Extended Data Fig.~\ref{fig:outflow}). 
In addition, the scenario in which multiple systems form by dynamical capture in a forming cluster, which have typical separations $>$10$^{3}$ au and significant different velocities \citep{Cournoyer-Cloutier2021}, is inconsistent  with the observed separation distributions and small velocity differences.  

If the parental disks are small and the separations of fragments are significantly widened within a short timescale \cite{Mignon-Risse2023}, we cannot completely rule out the possibility that disk fragmentation is responsible for the  close binary systems (e.g., C22-C38 and C35-C40). 
Overall, these results demonstrate that the majority of detected multiple systems is  formed from core fragmentation, although disk fragmentation may still occur on smaller scales than those we can resolve with the current spatial resolution.  
The measured separations of the multiple systems (a mean value of 731 au) are smaller than the expected Jeans lengths of 5000--10000~au for thermal Jeans fragmentation of the parent cores of the multiple systems (\hyperref[sec:methods]{Methods}). 
The estimated Jeans lengths are conservative upper limits to the separations of fragments since the derived volume densities of parental cores are lower limits due to the missing flux. 
If one assumes that the turbulence is acting as an isotropic support, the turbulent Jeans fragmentation yields even larger separation than that of the thermal one because the non-thermal velocity dispersion is higher than the thermal velocity dispersion. 
On the other hand, the anisotropic turbulent motions could promote collapse \citep{Ballesteros-Paredes2006}, in which the Jeans length could be reduced.  
The discrepancy indicates that the core fragmentation, which facilitates the formation of multiple system, is not merely regulated by the thermal pressure and/or turbulence, 
but additional mechanisms might also be important.  
For instance, ongoing global collapse and dynamical interactions of multiple systems could lead to inward migration \citep{Lee2019}, which moves the fragments closer together.

\begin{figure}[!h]
\centering
\includegraphics[width=1\textwidth]{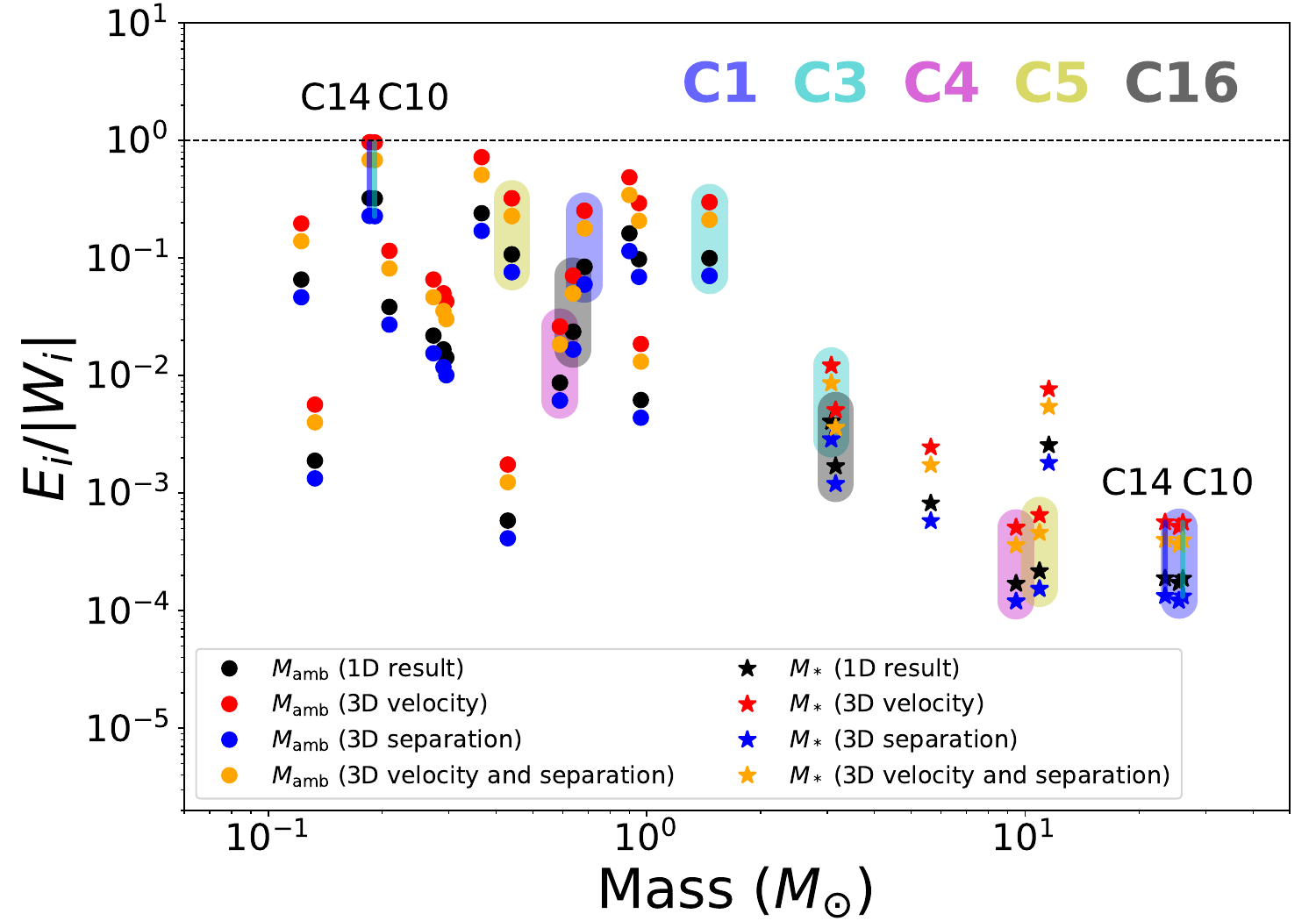}
\caption{
\textbf{The kinetic-to-gravitational energy ratio $E_i/|W_i|$ as a function of mass for the multiple  systems.}   The multiple systems with kinetic-to-gravitational energy ratio below unity are considered to be gravitationally bound. 
The circles and stars symbols are the results derived from ambient mass ($M_{\rm amb}$) and protostellar mass ($M_\ast$), respectively. 
$E_i/|W_i|$ has been estimated with four different methods:
(1) line-of-sight velocity difference and on-sky separation  (refer to one-dimensional, 1D; black symbols),
(2) three-dimensional (3D) velocity difference ($\sqrt{3}$ times the line-of-sight velocity difference) and on-sky separation (red symbols),
(3) line-of-sight velocity difference and 3D separation ($\sqrt{2}$ times the on-sky separation; blue symbols),
(3) 3D velocity difference and 3D separation (orange symbols). 
The black dashed line marks $E_i/|W_i|$ = 1. 
The members of quintuple system are marked with different color shadows, i.e., C1 (blue), C3 (cyan), C4 (red), C5 (yellow), and C16 (black).   
There are two condensations (C10 and C14) with $E_i/|W_i|\, >$ 1 for the 3D velocity scenario in the case of using $M_{\rm amb}$. 
If the central protostar mass is considered, the $E_i/|W_i|$ of these two condensations is smaller than 1. 
This figure indicates that all multiple systems are gravitationally bound. 
}
\label{fig:fig3}
\end{figure}

\vspace{0.5em}
\noindent{\textbf{Masses of the central protostars}}

Conventionally, the mass ($M_\ast$) of the central protostar can be estimated through modelling the rotation in a Keplerian disk. 
However, the estimation of dynamical masses of the central protostars is prevented by the non-detection of disk kinematic structures toward the multiple protostellar systems.    
On the other hand, a roughly $M_\ast$ can be estimated from the bolometric luminosity and a zero-age main sequence (ZAMS) assumption for the young protostars, which could provide a mass comparable to the dynamical mass obtained from Keplerian rotation \citep{Lu2022}. 
The luminosities of the central protostars are estimated from the temperature profile with the assumption that the spectral index of dust emissivity at far-infrared wavelengths is $\beta$ = 1 (\hyperref[sec:methods]{Methods}).
The derived luminosities range between 50.9 and 1.5$\times10^5$ \lsun, corresponding to B9 to O6.5 spectral type ZAMS stars \citep{panagia1973}. 
The masses are estimated to be between roughly 2.5 and 26.1 \msun\ according to the mass-luminosity relation \citep{Eker2018} (Table~\ref{tab1}). 
The derived masses are regarded as upper limits because the luminosities could be overestimated (\hyperref[sec:methods]{Methods}). 
Three binary systems (C22-C38, C39-C42, C34-C40), C12, and C29 do not have temperature measurements due to non-detection of CH$_3$CN and its isotopologues. 
There are at least 3 condensations (C1, C10, C14) with a mass of $>8$ \msun, even if the luminosity reaches a minimum value when $\beta$ decreases down to 0.  
This shows that indeed high-mass protostars exist in the quintuple and quadruple systems, which is consistent with the presence of Class II CH$_{3}$OH maser emission toward these regions (Fig.~\ref{fig:fig1}). 
The results indicate that high- and low-mass multiple protostellar  systems are simultaneously forming within G333.

\vspace{0.5em}
\noindent{\textbf{Stability of the multiple systems}}

To determine whether a multiple system is bound, we used the ambient mass $M_{\rm amb}$ to compute the kinetic ($E_i$) and gravitational ($W_i$) energies for each member of the multiple systems (\hyperref[sec:methods]{Methods}).   
The derived $E_i$ is smaller than $|W_i|$ for all condensations (Fig.~\ref{fig:fig3}), suggesting that these multiple systems are gravitationally bound, except for two condensations (C10 and C14). 
If including the central protostellar masses, the multiple systems would be even more gravitationally bound because the gravitational energy will be even higher. 
Indeed, the $E_i$ and $W_i$ computed from protostellar mass $M_\ast$ show that the $E_i$/$|W_i|$ ratio is below 0.1 and  smaller than those  derived from the ambient mass (Fig.~\ref{fig:fig3}). In addition, the $E_i$/$|W_i|$ ratio will be even smaller when both $M_{\rm amb}$ and $M_\ast$ are included. 
Therefore, the multiple systems are gravitationally bound at the present stage. 
With 20 identified multiple systems in G333, the observed multiplicity fraction is MF = 20/44 $\approx$ 45\%,  which is the fraction of systems that are
multiples (binary, triple, etc.), and companion frequency is CF = 46/44 $\approx$ 1.0, which is the average number of companions per system. 
The derived MF and CF are higher than those measured in Orion and Perseus star-forming regions for a similar separation range of 300--1400 au \citep{Tobin2022}, indicating that the multiplicity could be higher in denser cluster-forming environments.  
The estimated MF and CF are regarded as lower limits because further fragmentation might occur at smaller scales than what we can resolve with the current observations, and low-mass objects could be missed due to the limited sensitivity. 
The results indicate that the multiplicity in clusters is established in the protostellar phase.

\vspace{0.5em}
\noindent{\large \textbf{Perspectives}}

The discovery of these quintuple, quadruple, triple, and binary protostellar systems is the best observational evidence to show the imprints of core fragmentation in building multiplicity in high-mass cluster-forming environments. 
Although we cannot test if disk fragmentation is more important at smaller scales than what we have observed so far,  
we expect that more systems similar to G333 will be discovered given the high resolutions and high sensitivities of ALMA observations. 
The statistics of these systems will help to benchmark the relative contribution of core fragmentation to the population of multiple stars in high-mass star clusters. Their properties will determine the initial conditions of multiple system formation, as well as the dynamical evolution in a cluster environment.


\clearpage

\phantomsection
\label{sec:methods}
\noindent{\large \textbf{Methods}}

\noindent\textbf{High-mass star-forming region G333.23--0.06}

G333.23--0.06  is a typical high-mass star-forming region\citep{Foster2011,Jackson2013,Hoq2013,Stephens2015} at a distance of 5.2 kpc \citep{Whitaker2017} associated with Class II CH$_{3}$OH maser emission \citep{Caswell2011}, which can only be excited in high-density regions by strong radiation fields, making it exclusively found in high-mass star-forming regions \citep{Menten1991,Breen2013}.  
G333 have a mass reservoir of $\sim$3000 \msun{} with a mean column density of $1.6 \times 10^{23}$ cm$^{-2}$ within a 1.2 pc  radius \citep{Urquhart2018}. As a representative example of high-mass star-forming region, G333 provides an ideal laboratory to study the binary and high-order multiple systems formation in the cluster environment.

\noindent\textbf{Observations and data reduction}

Observations of G333 were performed with ALMA in Band 6 (at the wavelength of 1.3~mm) with the 12-m array using 41 antennas in configuration similar to C40-5 (hereafter short-baseline or low-resolution) on  November-05-2016 and 42 antennas in configuration similar to C43-8 (hereafter long-baseline) on July-28-2019 (Project ID: 2016.1.01036.S; PI: Sanhueza). Observations were obtained as part of the Digging into the Interior of Hot Cores with ALMA (DIHCA) project \citep{Olguin2022,Olguin2021,Taniguchi2023}. 
The baseline lengths are 18.6--1100~m and 91--8547~m for short-baseline and long-baseline observations, respectively. The correlators were tuned to cover four spectral windows with a spectral resolution of 976.6 KHz ($\sim$1.3 \kms) and a bandwidth of 1.875 GHz. These windows covered the frequency ranges of 233.5--235.5 GHz, 231.0--233.0 GHz, 219.0--221.0 GHz, and 216.9--218.7 GHz.   
The quasar J1427--4206 was used for flux and bandpass calibration, and J1603--4904 for phase calibration. The total on-source time toward the G333 is 6~minutes for short-baseline observations  and 19.6~minutes for long-baseline observations.   
The phase center used is ($\alpha$(ICRS), $\delta$(ICRS)) = 16h19m51.20s, $-$50\arcdeg{}15\arcmin{}13\farcs{00}.

The visibility data calibration was performed using the CASA  (version 5.4.0-70) software package \citep{McMullin2007}. We produced  continuum data from line-free channels and continuum-subtracted data cubes for each observation epoch using the procedure described in ref.\citep{Olguin2021}. We performed phase only self-calibration using the continuum data and the self-calibration solutions were applied to data cubes.  
To recover the extended emission, we combined the short-baseline and long-baseline self-calibration data for both continuum and data cubes (hereafter combined or high-resolution data). 
We produced images for short-baseline and combined data sets, separately.  
We used the TCLEAN task with Briggs weighting and a robust parameter of 0.5 to image the continuum. The resultant continuum images have a synthesized beam of 0.35\arcsec $\times$ 0.30\arcsec{} (1820 au$\times$1560 au, panel b of Fig.~\ref{fig:fig1}) with a position angle of  P.A. = -46.18\arcdeg{},  0.059\arcsec $\times$ 0.038\arcsec{} (307 au$\times$198 au) with a P.A. = 56.23\arcdeg{}, and 0.066\arcsec $\times$ 0.039\arcsec{} (343 au$\times$203 au, panels c--g of Fig.~\ref{fig:fig1}) with a P.A. = 54.47\arcdeg{} for short-baseline, long-baseline and combined dataset, respectively. The achieved 1$\sigma$ rms noise levels continuum images are about 0.16 mJy beam$^{-1}$,  0.05 mJy beam$^{-1}$, and 0.05 mJy beam$^{-1}$ for short-baseline, long-baseline, and combined data, respectively. 
Data cubes for each spectral window were produced using the automatic masking procedure YCLEAN \citep{Contreras2018}, which automatically cleans each map channel with custom-made masks. 
More details on the YCLEAN algorithm can be found in ref.\citep{Contreras2018}. 
The lines images 1$\sigma$ rms noise are  about 10~mJy beam$^{-1}$, 3~mJy beam$^{-1}$, and 3~mJy beam$^{-1}$ with a channel width of $\sim$0.65~\kms{} for short-baseline, long-baseline, and combined data, respectively. 
The largest recoverable angular scales are 3.5\arcsec{} for short-baseline and combined data, as determined by the short-baselines in the array. 

The Australia Telescope Compact Array (ATCA) 3.3~mm continuum image is retrieved from ref. \citep{Stephens2015} (panel a of Fig.~\ref{fig:fig1}). 
All images shown in this letter are prior to primary beam correction, while all measured fluxes have the primary beam correction applied.

\vspace{0.2em}
\noindent\textbf{Dense core and condensation identification}

To describe the dense molecular structures, we follow the nomenclature in the literature in which cores refer to structures with sizes of $\sim 10^{3}-10^{4}$ au, and condensations refer to substructures within a core. 

We use the \href{http://dendrograms.org/}{\tt astrodendro} algorithm and CASA-{\tt imfit} task to extract dense cores from short-baseline 1.3~mm continuum image   
and condensations from the combined and long-baseline only 1.3~mm continuum images. 
The {\tt astrodendro} identifies  the changing topology of the surfaces as a function of contour levels and extracts a series of hierarchical structure over a range of spatial scales \citep{Rosolowsky2008}. The performance of {\tt astrodendro} in characterizing the dense structure parameters (e.g., size and position angle) 
is not always good, while CASA-{\tt imfit} performs better in this regard via a two-dimensional Gaussian fit to the emission. 

Therefore, we used  {\tt astrodendro} to pre-select dense structures 
(i.e., the leaves in the terminology of  {\tt astrodendro}) 
from the 1.3~mm continuum images. We then use the parameters of the  pre-select structures from  {\tt astrodendro} as input to CASA-{\tt imfit} for more accurate measurement of their parameters, including peak position, peak flux ($I_{\rm peak}$), integrated flux ($S_{\nu}$),  major and minor axise sizes (full width at half maximum; FWHM$_{\rm maj}$ and FWHM$_{\rm min}$), and position angle (PA).

The following parameters are used in computing the dendrogram: the minimum pixel value $min_{-}value$ = 5$\sigma$, where $\sigma$ is the rms noise of the continuum image;  the minimum difference in the peak intensity between neighboring compact structures $min_{-}delta$ = 1$\sigma$;  the minimum number of pixels required for a structure to be considered an independent entity $min_{-}npix$ = $N$, where $N$ is the number of pixels in the synthesized beam area.

To remove suspicious condensations around the strong emission regimes caused by the diffuse emission in the combined data,  we have performed a cross-comparison of condensation catalogue derived from the combined data with the condensations revealed by the long-baseline only data.  
We identified 30 dense cores in short-baseline 1.3~mm continuum image and 44 condensations in combined 1.3~mm continuum image. Extended Data  Figs.~\ref{fig:lowcont} and  \ref{fig:highcont} show the identified dense cores and condensations, respectively (see also Table~\ref{tab1} and Extended Data Fig.~\ref{fig:separation} for the properties of multiple systems).

\vspace{0.2em}
\noindent\textbf{Centroid velocity of condensation}

The centroid velocity ($\rm V_{lsr}$) of each condensation is determined by gaussian fitting a CH$_{3}$OH $4_{2,2}-3_{1,2}$ ($E_{u}/k$ = 45.46 K) line that is detected in the majority of condensations in order to measure $\rm V_{lsr}$ in the same manner. 
The measured $\rm V_{lsr}$ have been validated by comparing with other dense gas tracers. 
We identify no clear velocity difference between the members of each multiple system (Table~\ref{tab1}), i.e., $\triangle \rm V_{lsr} < 1$ \kms{} that is smaller than the line-of-sight velocity differences (2.0--9.5  \kms{}) of the binary protostars in refs. \citep{Zhang2019,Tanaka2020}, 
indicating that all members of each multiple system are associated with the same region. 
We show the moment maps of CH$_{3}$OH in Extended Data Fig.~\ref{fig:ch3oh} for reference.

\vspace{0.2em}
\noindent\textbf{Jeans length}

If the fragmentation is governed by Jeans instability, the Jeans length $\lambda_{J}$, which is the separation between fragments, can be calculated by   \citep{Kippenhahn2013}
\begin{equation}\label{equ:sblaw_garay}
\lambda_{J} = \sqrt{\frac{\pi \, \sigma^{2}_{\rm eff}}{G\, \rho}} = 0.06\; {\rm pc} \, \left(\frac{\sigma_{\rm eff}}{0.188\; {\rm km\, s^{-1}}} \right) \left(\frac{\rho}{10^5 \; \rm cm^{-3}}\right), 
\end{equation}
where $\sigma_{\rm eff}$ is the effective velocity dispersion, $\rho$ is the volume density, and $G$ is the gravitational constant. 
The $\sigma_{\rm eff}$ equals the sound speed $c_{\rm s}$ for thermal Jeans fragmentation. The temperatures and volume densities of the parent cores used in Jeans analysis are $T$ = 80--340 K and $\rho$ = 3 $\times\, 10^{6}$--1 $\times\, 10^{7}$~cm$^{-3}$, respectively.  The derived thermal Jeans lengths range from 5000 to 10000 au.  
In the turbulent Jeans fragmentation scenario, the $\sigma_{\rm eff}$ includes  thermal and non-thermal velocity components, $\sigma_{\rm eff} = \sqrt{c_{\rm s}^2 + \sigma_{\rm nth}^2}$, under the assumption that the turbulence is acting as an isotropic of support.  The measured line widths  of \mthc{} are 3--5 \kms{} for the parent cores, resulting  turbulent Jeans lengths of 20000--54000 au.

\vspace{0.2em}
\noindent\textbf{Search for disk kinematic structure}

The observations cover the typical disk tracers, including \mthc{} and its isotopologues, and \mf{}, 
as well as other dense gas tracers, for instance H$_2$CO and its isotopologues, CH$_3$OH and its isotopologues, HC$_3$N, NH$_2$CHO, SO$_{2}$, SO, HNCO, HCOOH, $^{13}$CS, and OCS.   
Using these molecular lines, we have searched for disk-like rotating structures for the multiple systems and their parent cores with both short-baseline and combined data which have a channel width of $\sim$0.65 \kms. 
The dense gas tracers are not sufficiently strong to allow a reliable determination of kinematic information for 3 binary systems (C22-C38, C39-C42, C35-C40) in the combined data.

There are no obvious sign of disk kinematic structures toward the parent cores of multiple systems in any of the lines we examined based on short-baseline and combined data (Fig.~\ref{fig:fig2} and Extended Data Fig.~\ref{fig:lowmom1}).  
There are some lines with a velocity gradient in some dense cores, but no clear Keplerian disk-like rotating structure are found in the position-velocity (PV) diagram toward these cores. 
The velocity gradients trace either the outflows, or the large scale gas motions (e.g., gas flow, toroidal motions\citep{Beltran2016}).   
These dense cores are associated with unipolar, bipolar, and/or perpendicular outflows identified by the SiO emission from the ALMA short-baseline data (Extended Data Figs.~\ref{fig:lowmom1} and \ref{fig:outflow}).  
The detected misaligned outflows indicate that the embedded multiple systems do not come from the same co-rotating structures \citep{Lee2016}. This further suggests that the quintuple, quadruple, and triple systems are formed from core fragmentation \citep{Offner2016}.  
The detailed analysis of molecular outflows is beyond the scope of this letter, and will be presented in a future paper.

We examined the multiple systems following the same routine but using the combined data, and similarly found signs of velocity gradient in some condensations, but no obvious rotational signatures of disks (Fig.~\ref{fig:fig2} and  Extended Data Fig.~\ref{fig:highmom1}). 
Some velocity gradients are likely dominated by the outflows, while the others require higher angular resolution and sensitivity to spatially resolve the origin (e.g., unresolved outflows or accretion flows).  
At the early evolutionary stages of massive star formation, the size of disks might still be very small, potentially smaller than 100 au (refs.\citep{Kuiper2011,Oliva2020,Mignon-Risse2021}).  Considering the G333 is still at early evolutionary phases, the non-detection of disk structures indicates that the size and/or mass of disks are smaller than what we can resolve with the current observations. The current spatial resolution is  $\sim$260 au and 3$\sigma$ point source mass sensitivity of  $\sim$0.03 \msun{} assuming a temperature of 50 K.

As shown in Extended Data Fig.~\ref{fig:highmom1},  there is a redshifted velocity feature surrounding the blueshifted velocity toward C10 an C14. 
Two velocity components are detected toward C10 and C14. 
We have inspected these two velocity components separately, and found no obvious disk kinematic (Extended Data Fig.~\ref{fig:highmom11}).  
Several mechanisms could lead to two velocity components toward C10 and C14, such as unresolved multiple sources, unresolved Keplerian disk, or unresolved protostellar feedback within the condensation. 
Higher spatial and spectral resolution observations are required to distinguish these possibilities and determine the origin of these multiple velocity components.

\vspace{0.2em}
\noindent\textbf{Estimate of gas temperatures}

We derived the gas temperatures ($T_{\rm gas}$) using the $K$-ladder of \mthc{} $J=12\text{--}11$ and \tmthc{} $J=13\text{--}12$ transitions with the XCLASS package \citep{Moller2017}. 
The Markov Chain Monte Carlo tasks built in XCLASS was used to explore the parameter space during the fitting process. 
For the combined data, the signal-to-noise ratios of the \mthc{} and \tmthc{} are not sufficient to derive a reliable temperature map in the majority of cores, and they are not detected toward  C12, C29, and three binary systems (C22-C38, C39-C42, and C35-C40). 
To improve the signal-to-noise ratio with minimal nearby source(s) contamination, we averaged the spectra within a half beam size toward the condensations.  
We exclude the $K < 4$ ladders in regions where these lower energy transitions become optically thick, i.e., where line profile show self absorption or saturated emission. 
Although optical depth and chemical processes might affect the estimated temperatures, this is the best estimate of the temperature structure of the inner envelope we can obtain. 
To improve the fitting of the \mthc{} and \tmthc{} lines, we include  CH$_{3}^{13}$CH $J=12\text{--}11$ lines and other molecular lines (i.e., CH$_3$OH, HNCO) in fitting for \mthc{} and CH$_3$OCH$_3$ for 
\tmthc{}, if they are detected. 
The derived rotational temperatures range from 108 to 532 K (Table~\ref{tab1}).

The rotational temperatures derived from \tmthc{} are higher than those of \mthc{}. 
This is because the \mthc{} lines have a higher optical depth and preferably trace the surface of the structure, while  \tmthc{} is optically thinner and better trace the interior of the structure. 
This clearly suggests that these objects exhibit temperatures gradients and are internally heated by the protostar(s) at the centre. 
Therefore, we use the temperature derived from \tmthc{} to estimate the mass and luminosity, and in the case that \tmthc{} is not sufficiently strong to allow a temperature measurement, we use the temperature derived from \mthc{}.  
Examples of spectra of  \tmthc{} and \mthc{} for the quadruple system are presented in Supplementary Fig.~\ref{fig:spectra}.

\vspace{0.2em}
\noindent\textbf{Computing the luminosity of the embedded protostar}

With the derived gas temperature and taking into account the dust emissivity, we are able to approximately estimate the luminosity of the central heating source according to the relation between the  temperature distribution and embedded protostellar luminosity  \citep{scoville1976,garay1999}, which is given by following equation
\begin{equation}\label{equ:sblaw_garay}
L= 10^5~\lsun \, \times  \, \left(\frac{T_D}{65~\text{K}}\right)^{4+\beta} \left(\frac{0.1}{f}\right)^{-1}\left(\frac{0.1~\text{pc}}{r}\right)^{-2},
\end{equation}
where $T_D$ is the dust temperature at the radius $r$, $\beta$ is the spectral index of dust emissivity at far-infrared wavelengths and $f$ is its value at the wavelength of 50~\micron{}. 
The $\beta$ usually ranges from 0 to 1 
and $f$  is usually adopted as 0.1 (refs.\citep{Ossenkopf1994,garay1999}).  
The gas temperature derived from either \tmthc{}  or \mthc{}  based on the averaged spectrum within a half of beam size of the condensation's continuum peak can be used as a good approximation of $T_D$ at 
the radius $r$ = 130 au (corresponding to the half beam size of $\sim$0.025\arcsec), where the densities are sufficiently high ($> 10^{4.5} \, {\rm cm}^{-3}$) for the dust and gas to be well coupled  \citep{Goldsmith2001}. 
The \tmthc{} and \mthc{} lines are not detected in C12, C29, and 3 binary systems (C22--C38, C39--C42, C35--C40). Thus, we refrain from estimating the $M_\ast$ for these condensations to avoid the large uncertainty.

Assuming $\beta=1$, the derived luminosities range from 50.9 to 1.5$\times 10^{5}$ \lsun{} (Table~\ref{tab1}), corresponding to spectral  B9- to O6.5-type ZAMS stars \citep{panagia1973}, whose mass would be about 2.5 to 26.1~\msun{} according to the mass-luminosity (M-L) relation  \citep{Eker2018}. 
The total derived luminosity ($\sim$4.2$\times 10^{5}$ \lsun{}) is higher than the value ($\sim$2$\times 10^{4}$ \lsun{}) estimated on clump scale that has an uncertainty up to a factor of a few \citep{Urquhart2018}.  
We note that the derived temperatures could have some contaminations from neighbouring sources, and the luminosity of clump should be treated as a lower limit of total bolometric luminosity due to the lack of robust measurement at mid-infrared \citep{Urquhart2018}. 
The derived masses are regarded as conservative upper limits. 
There are 5 condensations (C1, C5, C8, C10, and C14) with estimated luminosities of 8.0$\times 10^{3}$ -- 1.5$\times 10^{5}$ \lsun, corresponding to a B1--O6.5 spectral type ZAMS star of $>$8 \msun{} among the multiple systems (Table~\ref{tab1}). 
The derived luminosity will be lower if a smaller $\beta$  is adopted. 
However, even if $\beta$=0 is adopted, there are still 3 condensations (C1, C10, and C14) associated with a $M_\ast \, >$8 \msun{}. 
Therefore, a massive protostar should exist in both quintuple and quadruple systems, as also suggested by the presence of Class II CH$_{3}$OH maser, which are excited in high-density regions by strong radiation fields and exclusively tracing high-mass star-forming regions (Fig.~\ref{fig:fig1}).

\vspace{0.2em}
\noindent\textbf{Estimating ambient gas mass from dust continuum emission}

The brightness temperatures of the dust emission in the condensations are lower than the gas temperatures $T_{\rm gas}$ which is a good approximation of $T_{D}$. 
To check if the dust emission is optically thin, we computed the  optical depth $\tau_{\rm cont}$ of the continuum emission at the peak position of each condensation using \citep{Frau2010}
\begin{equation}
\label{equ:tau}
\tau^{\rm beam}_{\rm \nu} = -{\rm ln}\left(1 \, - \, \frac{S^{\rm beam}_{\rm \nu}}{\Omega_{\rm A} B_{\nu}(T_D)} \right)
\end{equation} 
where $B_{\nu}$ is the Planck function at the dust temperature $T_D$, $S^{\rm beam}_{\rm \nu}$ is the continuum peak flux density, $\Omega_{\rm A}$ is the beam solid angle. 
The condensations are dense enough ($>$10$^{4.5}$~\cc{}) for gas and dust to be well coupled and in thermal equilibrium. As such, the gas temperature derived from the \tmthc{}  or \mthc{} should be approximately equal to the dust temperature.  
The derived optical depths are 0.04--0.27 with a mean value of 0.1 for all available condensations, indicating optically thin dust emission.

The observed 1.3~mm continuum emission is dominated by  thermal dust emission toward G333 because the hydrogen recombination line (i.e., H30$\alpha$) is not detected toward condensations and the ATCA 3.3~mm continuum emission is also dominated by dust emission \citep{Stephens2015}. 
We calculate the ambient gas mass for the condensations following 
\begin{equation}
\label{Mgas}
M_{\rm amb} = \eta \frac{S_{\nu} \, \rm{D^2}}{\kappa_{\nu} \, B_{\nu}(T_D)},
\end{equation}
where $\eta$ = 100 is the gas-to-dust ratio, 
$S_{\nu}$ is the measured integrated source flux,  
 $m_{\rm H}$ is 
the mass of an hydrogen atom, $\mu$ = 2.8 is the mean molecular 
weight of the interstellar medium,  
D = 5.2 kpc is the distance to the source, 
and $\kappa_{\nu}$ is the dust opacity at a frequency of $\nu$. 
We adopted a value of 0.9 cm$^{-2}$ g$^{-1}$ for 
$\kappa_{\rm 1.3 mm}$, which corresponds to the opacity 
of thin ice mantles and a gas density of 10$^{6}$ cm$^{-3}$ 
(ref. \citep{OH1994}). 
We use the lowest temperature of 108 K derived from \mthc{} as an approximation to the temperature for the condensations in which \tmthc{} and \mthc{} are not detected. The actual temperature should be lower than 108 K, indicating that the derived mass is the lower limit for the condensations. 
The derived ambient gas masses of multiple systems are between 0.19 and 1.47 \msun{} (Table~\ref{tab1}), with the mean and median values of 0.59 and 0.52 \msun, respectively.  
The estimated ambient gas masses should be regarded as lower limits due to the interferometric observations suffering from missing flux.

\vspace{0.2em}
\noindent\textbf{Stability analysis of multiple system}

To assess the stability of the multiple system, we compute the potential and kinetic energy of each object  following the approach introduced in ref.\cite{Pineda2015}. 
The gravitational potential energy, $W_{i}$, and kinetic energy, $E_{i}$, can be caculated by  
\begin{equation}
\label{equ:grav}
W_{i}  = - \sum_{i \neq j}  \frac{ G m_{i} m_{j} }{ r_{ij} },
\end{equation} 

\begin{equation}
\label{equ:kine}
E_{i}  = \frac{1}{2} m_{i} (\rm V_{i} - V_{com})^2,
\end{equation} 
where $m_{i}$ and $m_{j}$ are the masses of object $i$ and $j$, $r_{ij}$ is the separation between $i$ and $j$, V$_{i}$ is the (line-of-sight) velocity of object $i$, and V$_{\rm com}$ is the velocity of the centre of mass of the system. We determine the V$_{\rm com}$ through 
\begin{equation}
\label{equ:grav}
{\rm V_{com} }  =  \frac{\sum\limits_{k} m_{k} V_{k}}  {\sum\limits_{k} m_{k}} ,
\end{equation}
where  $m_{k}$ and V$_{k}$ are the mass and velocity of the object $k$ in the multiple system. 
A star with $E_{i}/|W_{i}| \,<$ 1 is considered to be bound to the system.

The full velocity difference is $\sqrt{3}$ times the velocity difference along the line-of-sight, $\triangle \rm V_{\rm 3D}$ = $\sqrt{3} \triangle \rm V_{\rm 1D}$ = $\sqrt{3}$($\rm V_{i} - V_{\rm com}$), assuming the measured velocity difference is representative of the one-dimensional velocity difference.  
Similarly, the total separation is $\sqrt{2}$ times the projected separation on the sky, $r_{\rm 3D}$ = $\sqrt{2} r_{\rm 1D}$ = $\sqrt{2} r_{ij}$, assuming that the measured projected separation is a good  approximation of the separation along the line-of-sight.  
The observed mean separation, $\langle r_{\rm 1D}\rangle$, is about 731 au, which is consistent with the typical projected value of 700~au, $r_{\rm 1D}$ = $r_{\rm 3D}/\sqrt{2}$ = 1000/$\sqrt{2}$ = 700 au, in the simulation of multiple star formation via core fragmentation \citep{Kuruwita2023}.

We used the ambient mass $M_{\rm amb}$ and protostar mass $M_\ast$ to calculate the kinetic and gravitational energies for condensations in both one- and three-dimensional scenarios.  
We find that all multiple systems are gravitationally bound (Fig.~\ref{fig:fig3}), with exceptions for two condensations (C10 and C14) that have  $E_{i}/|W_{i}| \,>$ 1 for 3D velocity difference in the case of using $M_{\rm amb}$. 
However, these two condensations are gravitationally bound if the central protostar mass is considered (Fig.~\ref{fig:fig3}). 
If the total mass, $M_{\rm tot}$ = $M_{\rm amb}$ + $M_\ast$, is used, the $E_{i}/|W_{i}|$ ratio will be smaller. 
Therefore, we conclude that all multiple systems are consistent with being gravitationally bound at the present stage. 
Table~\ref{tab1} presents the $E_{i}$ and $W_{i}$ for each condensation.


\vspace{1em}
\noindent\textbf{Data Availability}

This paper makes use of the following ALMA data: ADS/JAO.ALMA\#2016.1.01036.S. The data are available at \url{https://almascience.nao.ac.jp/aq} by setting the observation code. 
Owing to their size, the reduced data used for this study are available from the corresponding authors upon reasonable request.


\vspace{0.2em}
\noindent\textbf{Code Availability}

The ALMA data were reduced using CASA versions 5.4.0-70 that are available at \url{https://casa.nrao.edu/casa_obtaining.shtml}. 
The {\tt pvextractor} package was used to make position velocity diagram, which  is available at \url{https://pvextractor.readthedocs.io}. 
The source identification package of {\tt astrodendro} is available at \url{http://dendrograms.org/}. The XCLASS is available at \url{https://xclass.astro.uni-koeln.de}.


\vspace{0.2em}
\noindent\textbf{Acknowledgements}

We thank Chang Won Lee, Kee-Tae Kim, Fr\'ed\'erique Motte, and Tom Megeath  for helpful discussions. 
P.S. was partially supported by a Grant-in-Aid for Scientific Research (KAKENHI Number JP22H01271 and JP23H01221) of JSPS. R.K. acknowledges financial support via the Heisenberg Research Grant funded by the German Research Foundation (DFG) under grant no.~KU 2849/9.  
R.E.P is supported by a Discovery grant from NSERC Canada. 
X.L acknowledges support from the National Key R\&D Program of China (No.\ 2022YFA1603101), Natural Science Foundation of Shanghai (No.\ 23ZR1482100), the National Natural Science Foundation of China (NSFC) through grant Nos.\ 12273090 and 12322305, and the Chinese Academy of Sciences (CAS) ``Light of West China'' Program (No.\ xbzgzdsys-202212). 
F.L is supported by the National Natural Science Foundation of China grant 12103024 and the fellowship of China Postdoctoral Science Foundation 2021M691531.  
This work made use of the following ALMA data: ADS/JAO.ALMA\#2016.1.01036.S.  ALMA is a partnership of ESO (representing its member states), NSF (USA) and NINS (Japan), together with NRC (Canada), MOST and ASIAA (Taiwan), and KASI (Republic of Korea), in cooperation with the Republic of Chile. The Joint ALMA Observatory is operated by ESO, AUI/NRAO and NAOJ. Data analysis was in part carried out on the Multi-wavelength Data Analysis System operated by the Astronomy Data Center (ADC), National Astronomical Observatory of Japan.


\vspace{0.2em}
\noindent\textbf{Author contributions}

S.L. led the data reduction, data analysis and interpretation of the data, writing major sections of the main text and Methods. 
P.S. led the ALMA proposal, and contributed to the interpretation of results and writing. 
H.B. assisted with the analysis and contributed to the interpretation of results. 
H.R.V.C. and R.K. contributed to the interpretation of the results. 
F.A.O. conducted the data calibration and commented on the article. 
R.E.P., I.W.S., Q.Z., F.N. and X.L. commented on and helped to improve the article. 
R.L.K. contributed to the interpretation of the separation of multiple systems. T.S., T.H, K.T. and F.L. commented on and helped to improve the article. 
All authors contributed to the discussed of the results and helped with manuscript preparation.


\vspace{0.2em}
\noindent\textbf{Competing interests}

The authors declare no competing interests.


\clearpage
\begin{sidewaystable}
\caption{Properties of condensations}\label{tab1}
\setlength\tabcolsep{2pt} 
\resizebox{1\textwidth}{!}{%
\scriptsize 
\begin{tabular*}{\textheight}{@{\extracolsep\fill}lcccccccccccccccc}
\toprule%
Condensation 	 & RA 		 &  DEC & $I_{\rm peak}$ & $S_{\nu}$ &  size	 & Radius & V$_{\rm lsr}$	 & 
$T_{\rm gas}$ 	 & $M_{\rm amb}$ 	 & $N_{\rm H_2}$&  $W_{i}(M_{\rm amb})$ &  $E_{i}(M_{\rm amb})$ &  $L_{\rm bol}$ &  $M_\ast$  &  $W_{i}(M_\ast)$ &  $E_{i}(M_\ast)$  \\
	 & (hh:mm:ss.sss) & (dd:mm:ss.sss) & (mJy beam$^{-1}$) & (mJy) &  (\arcsec$\times$\arcsec, deg) 	 & (\arcsec [au]) & (km s$^{-1}$)	 & 
(K) 	 & ($M_\odot$) 	 & (cm$^{-2}$)  &  ($M_\odot \rm \;km^{2} s^{-2}$) &  ($M_\odot \rm \;km^{2} s^{-2}$) 	 &  ($L_\odot$) &  ($M_\odot$)  &  ($M_\odot \rm \;km^{2} s^{-2}$) &  ($M_\odot \rm \;km^{2} s^{-2}$) \\
	 &  &  & & & & & & 
 	 & & ($\times 10^{24}$)  &  ($\times 10^{-2}$) &  ($\times 10^{-2}$) 	 &  & &  ($\times 10^{-2}$) &  ($\times 10^{-2}$) \\
\midrule
(1) & (2) & (3) & (4) & (5) & (6) & (7) & (8) & (9) & (10) & (11) & (12) & (13) & (14) & (15) & (16) & (17)	\\ \hline
C1 & 16:19:51.275 & -50:15:14.529 & 6.01  & 38.80$\pm$2.07  & 0.15$\times$0.09, 5  & 0.12 [607] & -84.15$\pm$0.21  & 525$\pm$66  & 0.68  & 2.50  & -138.03  & 12.91  & 136191  & 25.49  & -743.44  & 12.84 	\\
C3 & 16:19:51.271 & -50:15:14.760 & 2.58  & 19.21$\pm$0.94  & 0.15$\times$0.11, 87  & 0.13 [671] & -84.76$\pm$0.09  & 126$\pm$1  & 1.47  & 4.64  & -132.44  & 13.75  & 107  & 3.07  & -96.10  & 39.08 	\\
C4 & 16:19:51.290 & -50:15:14.548 & 3.27  & 18.79$\pm$2.00  & 0.13$\times$0.09, 17  & 0.11 [572] & -84.39$\pm$0.25  & 297$\pm$10  & 0.59  & 2.42  & -87.91  & 0.84  & 7973  & 9.47  & -525.32  & 8.95 	\\
C5 & 16:19:51.291 & -50:15:14.458 & 3.01  & 15.24$\pm$1.82  & 0.14$\times$0.08, 76  & 0.10 [532] & -84.11$\pm$0.28  & 323$\pm$14  & 0.44  & 2.05  & -85.09  & 10.56  & 11988  & 10.91  & -533.26  & 11.60 	\\
C16 & 16:19:51.269 & -50:15:14.410 & 1.74  & 8.57$\pm$0.65  & 0.11$\times$0.09, 110  & 0.10 [515] & -84.67$\pm$0.14  & 128$\pm$1  & 0.64  & 3.06  & -101.27  & 2.36  & 119  & 3.16  & -157.02  & 26.67 	\\ \hline

C6 & 16:19:50.938 & -50:15:10.020 & 2.99  & 14.21$\pm$0.74  & 0.11$\times$0.08, 141  & 0.09 [494] & -90.52$\pm$0.15  & 150$\pm$0  & 0.90  & 4.47  & -82.34  & 13.37  & 260  & 3.85  & & \\
C12 & 16:19:50.955 & -50:15:10.080 & 1.43  & 10.69$\pm$0.80  & 0.18$\times$0.09, 48  & 0.13 [678] & -89.95$\pm$0.13  & & 0.95  & 3.00  & -93.45  & 9.15  &   & & & \\
C26 & 16:19:50.952 & -50:15:09.980 & 1.65  & 3.95$\pm$0.58  & 0.07$\times$0.05, 30  & 0.06 [310] & -90.06$\pm$0.19  & 130$\pm$5  & 0.29  & 2.85  & -54.46  & 0.91  & 128  & 3.22  & & \\ \hline

C8 & 16:19:50.860 & -50:15:10.465 & 3.48  & 13.12$\pm$0.46  & 0.10$\times$0.07, 83  & 0.08 [438] & -88.05$\pm$0.12  & 333$\pm$0  & 0.37  & 2.29  & -17.88  & 4.04  & 14113  & 11.55  & -517.98  & 132.87 	\\
C10 & 16:19:50.884 & -50:15:10.520 & 4.52  & 10.99$\pm$0.69  & 0.06$\times$0.06, 62  & 0.06 [313] & -88.67$\pm$0.15  & 532$\pm$5  & 0.19  & 1.85  & -8.71  & 3.46  & 145836  & 26.10  & -1330.96  & 25.01 	\\
C14 & 16:19:50.878 & -50:15:10.604 & 2.63  & 9.97$\pm$0.81  & 0.10$\times$0.06, 131  & 0.08 [414] & -88.68$\pm$0.14  & 499$\pm$2  & 0.19  & 1.15  & -7.36  & 2.61  & 106984  & 23.43  & -1313.72  & 24.83 	\\
C17 & 16:19:50.850 & -50:15:10.397 & 1.45  & 6.37$\pm$0.48  & 0.11$\times$0.08, 25  & 0.09 [484] & -88.27$\pm$0.09  & 203$\pm$2  & 0.30  & 1.59  & -15.07  & 0.13  & 1178  & 5.63  & -238.09  & 19.55 	\\ \hline

C11 & 16:19:51.245 & -50:15:14.300 & 2.67  & 10.84$\pm$0.87  & 0.13$\times$0.05, 101  & 0.08 [427] & -86.11$\pm$0.17  & 108$\pm$2  & 0.97  & 5.61  & -27.12  & 0.17  & 51  & 2.47  & & \\
C29 & 16:19:51.235 & -50:15:14.241 & 1.33  & 3.06$\pm$0.52  & 0.09$\times$0.04, 55  & 0.06 [300] & -85.95$\pm$0.19  &   & 0.27  & 2.81  & -27.12  & 0.59  &   & & & \\ \hline

C22 & 16:19:51.241 & -50:15:12.540 & 1.21  & 4.81$\pm$0.43  & 0.10$\times$0.07, 144  & 0.08 [440] & -87.53$\pm$0.06  & & 0.43  & 2.56  & -7.47  & &   & & & \\
C38 & 16:19:51.235 & -50:15:12.610 & 0.73  & 1.49$\pm$0.33  & 0.09$\times$0.06, 18\footnotemark[1]  & 0.07 [364]\footnotemark[1] & -87.57$\pm$0.09  & & 0.13  & 1.54  & -7.47  & 0.01  &   & & & \\ \hline

C35 & 16:19:51.386 & -50:15:14.821 & 0.65  & 2.34$\pm$0.15  & 0.13$\times$0.05, 69  & 0.08 [417] & -85.04$\pm$0.12  & & 0.21  & 1.38  & -4.91  & 0.19  &   & & & \\ 
C40 & 16:19:51.389 & -50:15:14.878 & 0.61  & 1.37$\pm$0.12  & 0.10$\times$0.03, 72  & 0.05 [269] & -85.25$\pm$0.13  & & 0.12  & 1.28  & -4.91  & 0.32  &   & & & \\ \hline

C39 & 16:19:51.784 & -50:15:12.623 & 0.44  & 1.43$\pm$0.09  & 0.09$\times$0.06, 104  & 0.07 [387] & & & 0.13  & 0.92  & -0.37  & &   & & & \\
C42 & 16:19:51.793 & -50:15:12.879 & 0.53  & 0.73$\pm$0.10  & 0.05$\times$0.02, 69  & 0.03 [153] & & & 0.06  & 1.12  & -0.37  & &   & & & \\ \hline
\end{tabular*}}
\footnotetext{Note: Columns (1)--(3) present the name, right ascension, and declination of condensations.
The continuum peak intensity, flux density, beam-deconvolved size, beam-deconvolved radius, and centroid velocity are shown in columns (4)--(8).  
Column (9) is the gas temperature derived from  \tmthc{} or \mthc{}.  Column (10) and (11) are the ambient mass and the H$_2$ column density.  
The gravitation potential energy and kinetic energy derived from $M_{\rm amb}$ are shown in columns (12) and (13). 
The luminosity and corresponding mass of central protostar are presented in columns (14) and (15). 
Columns (16) and (17) shows the gravitation potential energy and kinetic energy derived from $M_\ast$. 
}
\footnotetext[1]{$^{1}$This is beam-convolved size due to the condensation marginally resolved.}
\end{sidewaystable}

\clearpage

\clearpage

\setcounter{figure}{0}
\renewcommand{\figurename}{Extended Data Fig.}
\renewcommand{\figureautorefname}{Extended Data Fig.}


\begin{figure}[!h]
\centering
\includegraphics[width=1\textwidth]{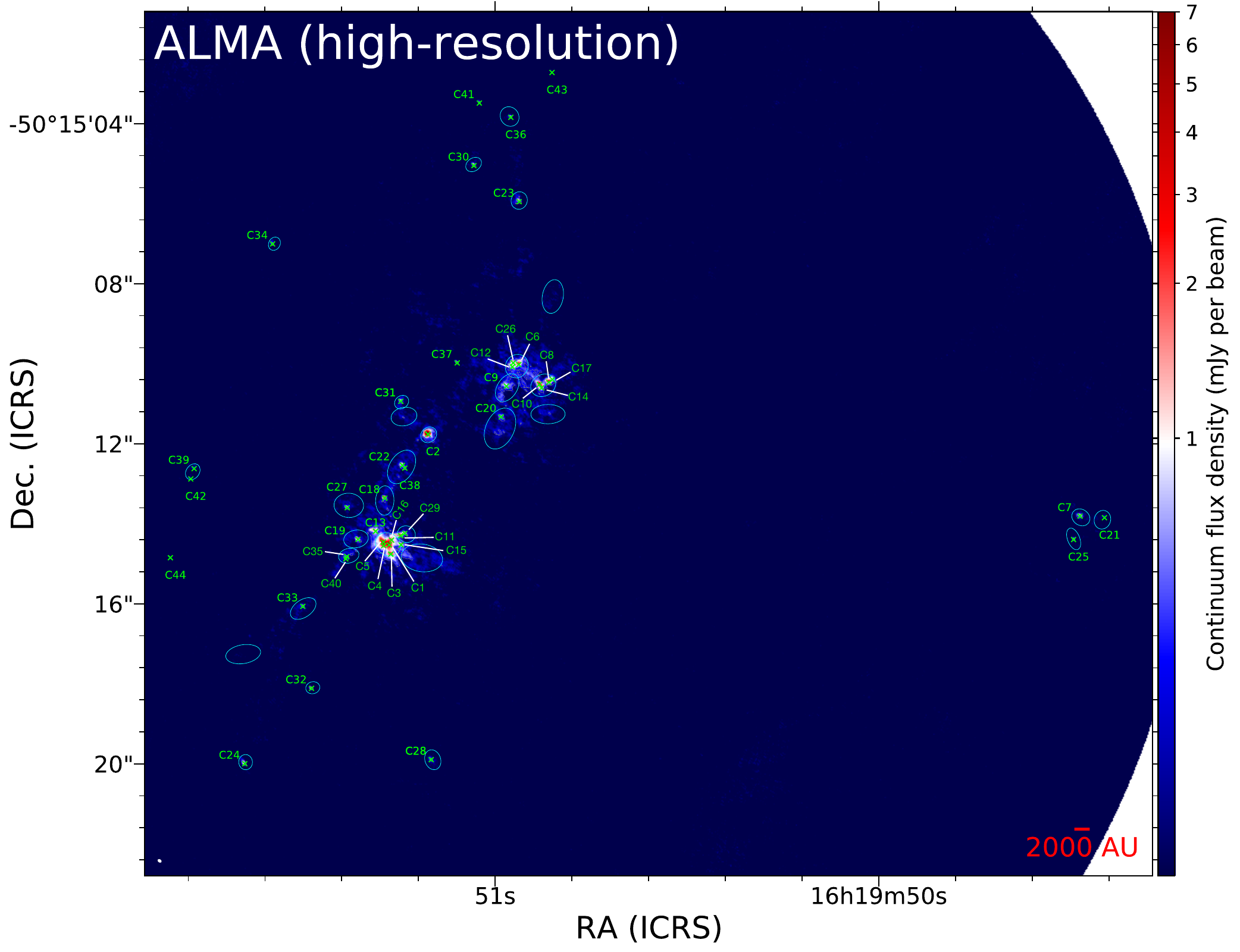}
\caption{
\textbf{ALMA high-resolution 1.3~mm continuum image.} 
The cyan ellipses are the identified dense cores as shown in Extended Data Fig.~\ref{fig:lowcont}.    
The green crosses show condensations identified from ALMA high-resolution 1.3~mm continuum image.  
The grey contour shows the 7$\sigma$, where $\sigma$ = 0.05 mJy beam$^{-1}$. 
The synthesized beam size of 1.3 mm continuum image present in the lower left corner with a white ellipse. 
}
\label{fig:highcont}
\end{figure}

\begin{figure}[!h]
\centering
\includegraphics[width=0.7\textwidth]{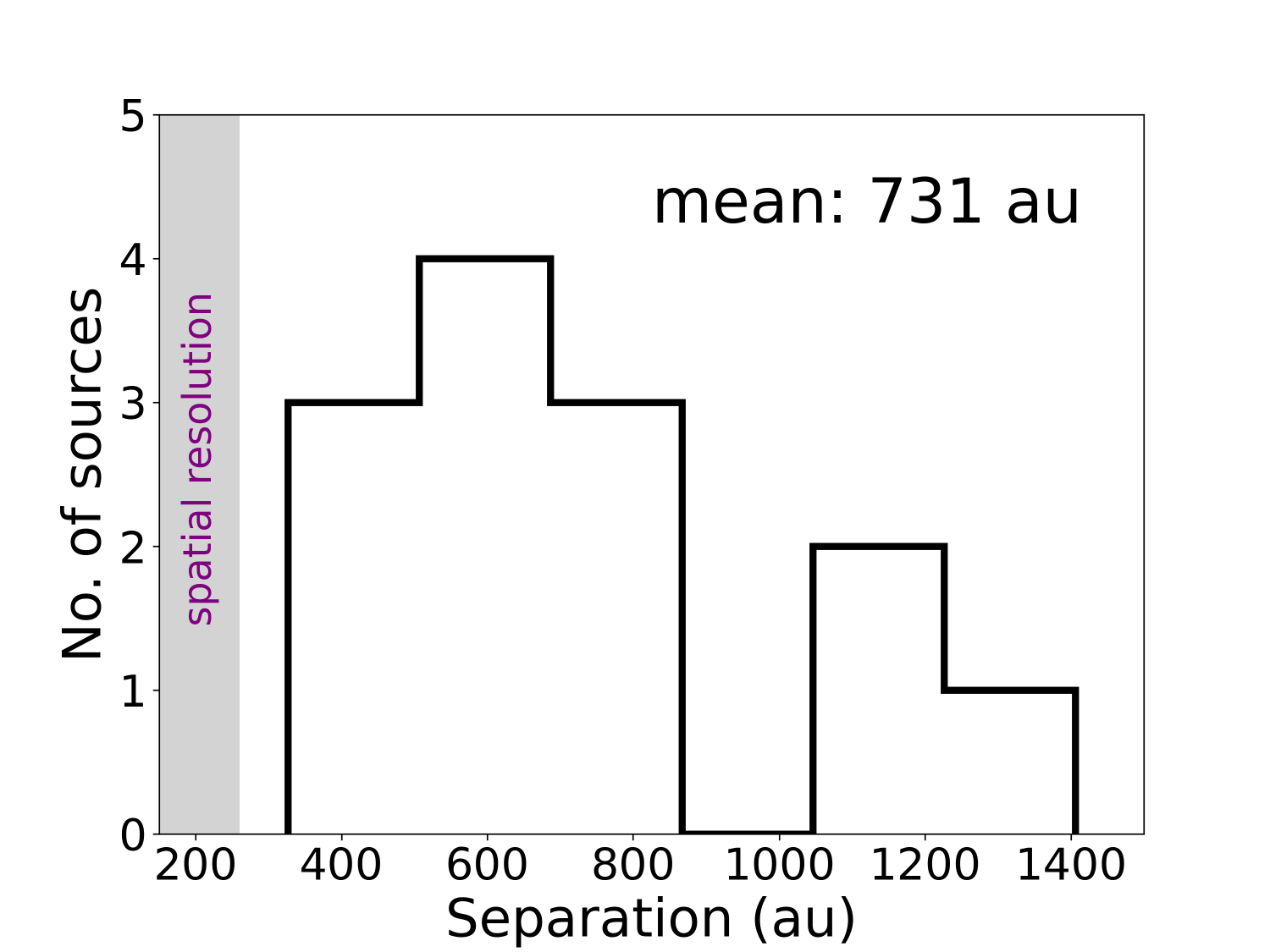}
\caption{
\textbf{The separation distributions of multiple systems.} 
The separations range from 327 to 1406 au, with a mean value of 731 au. 
}
\label{fig:separation}
\end{figure}

\begin{figure}[!h]
\centering
\includegraphics[width=1\textwidth]{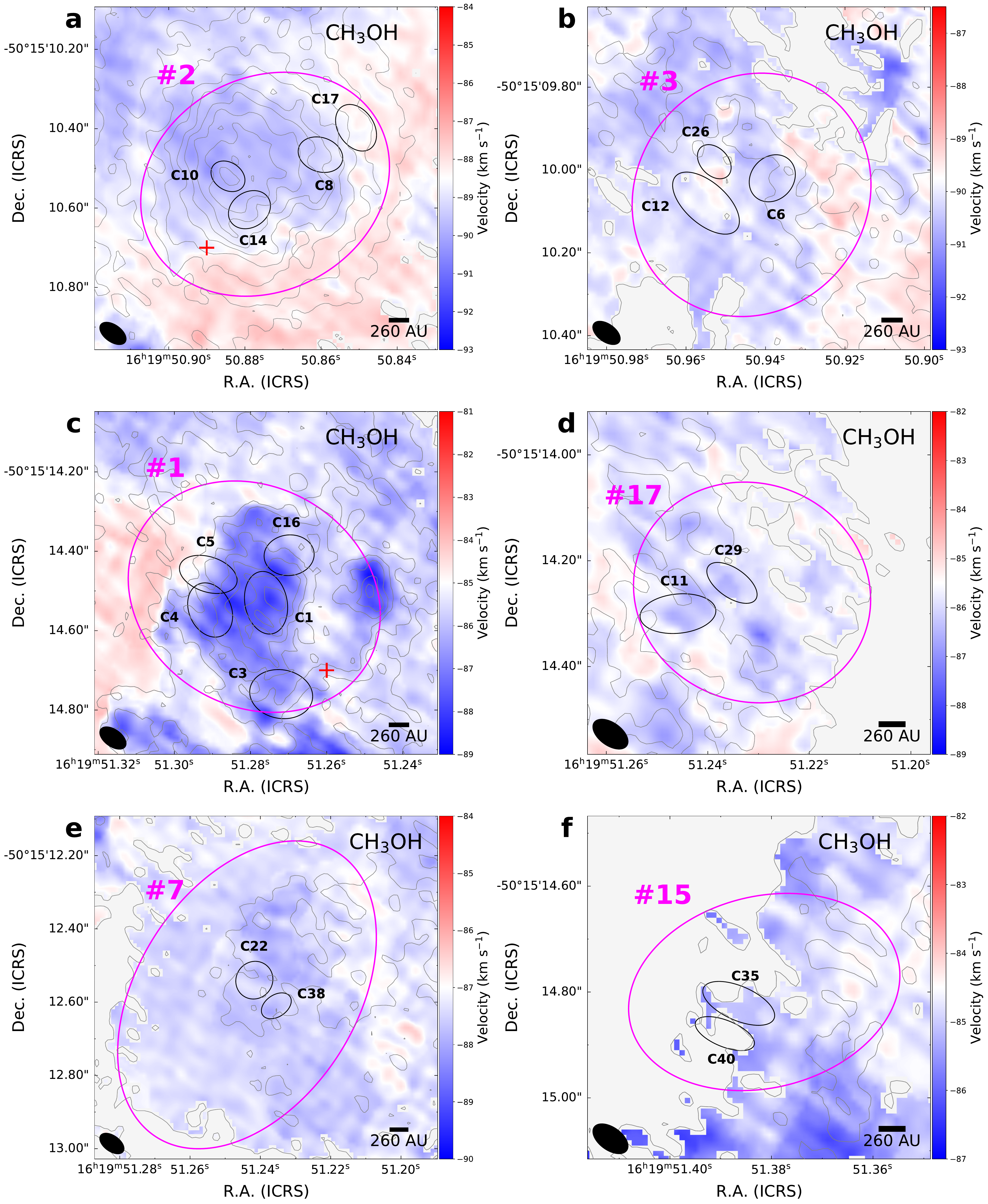}
\caption{
\textbf{Intensity-weighted velocity maps of CH$_{3}$OH derived from the high-resolution data toward multiple systems.} 
The grey contours show the corresponding velocity-integrated intensity maps. Contours levels start at 3$\sigma_{\rm rms}$ and increase in step of 3$\sigma_{\rm rms}$ interval, where $\sigma_{\rm rms}$ is 7.6,  6, 6.4, 6.4, 6, and 5 mJy beam$^{-1}$ km s$^{-1}$ for \textbf{a--f}. 
The condensation C40 does not have sufficient signal-to-noise (SNR) ratio for the moment maps, while the condensation averaged spectrum has sufficient SNR to determine the centroid velocity. 
The black ellipses in the lower left corner of each panel denote the synthesized beam of lines images. 
}
\label{fig:ch3oh}
\end{figure}

\begin{figure}[!h]
\centering
\includegraphics[width=1\textwidth]{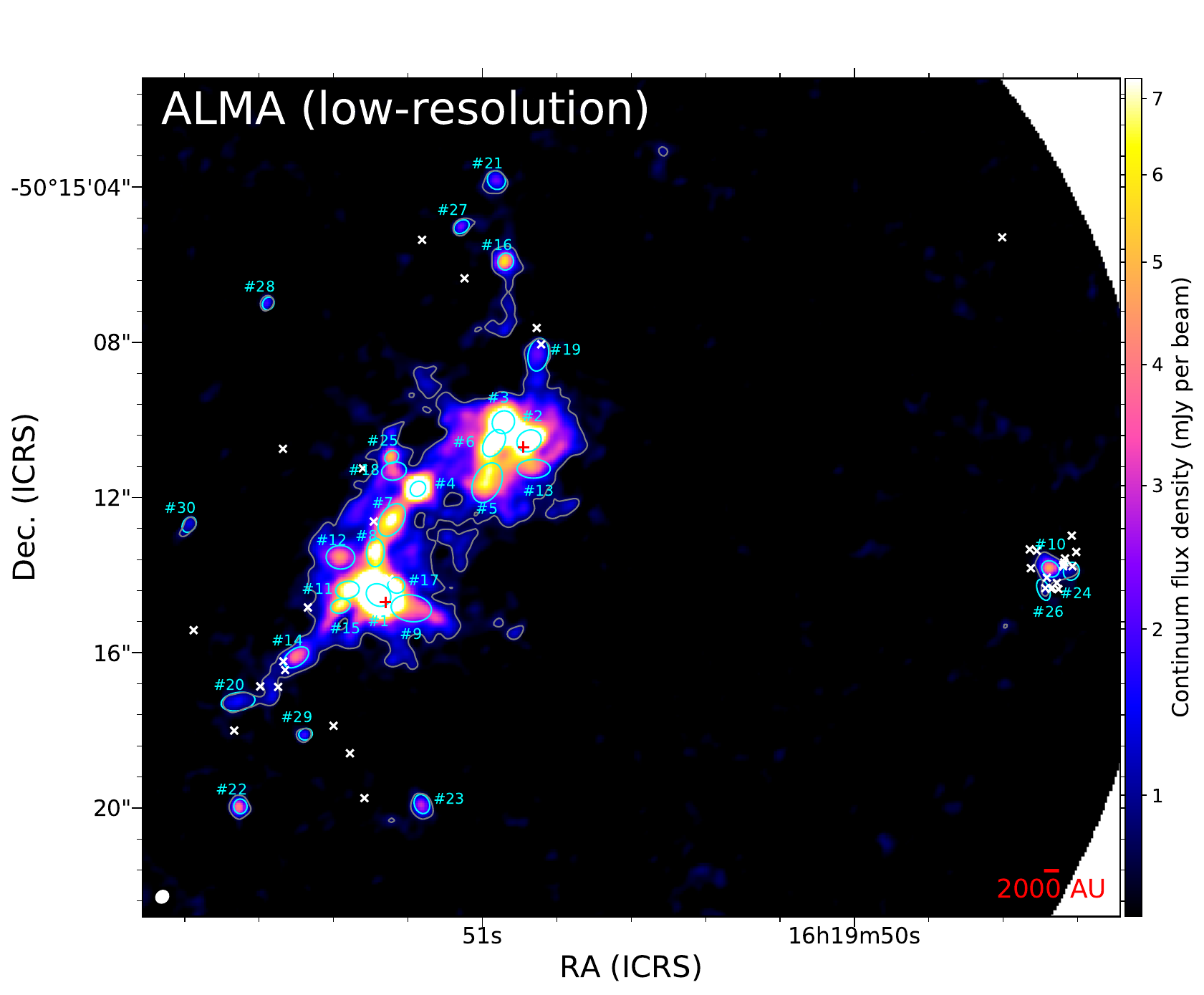}
\caption{
\textbf{ALMA low-resolution 1.3~mm continuum image.} 
The cyan ellipses are dense cores identified from ALMA low-resolution 1.3~mm continuum image.   
The grey contour shows the 7$\sigma$, where $\sigma$ = 0.16 mJy beam$^{-1}$. 
The red plus and white cross symbols are Class II (ref.~\citep{Caswell2011}) and Class I  (ref.~\citep{Voronkov2014}) CH$_3$OH maser, respectively, indicating intense ongoing star formation activity. 
The synthesized beam size of 1.3 mm continuum image present in the lower left corner with a white ellipse. 
}
\label{fig:lowcont}
\end{figure}

\begin{figure}[!h]
\centering
\includegraphics[width=1\textwidth]{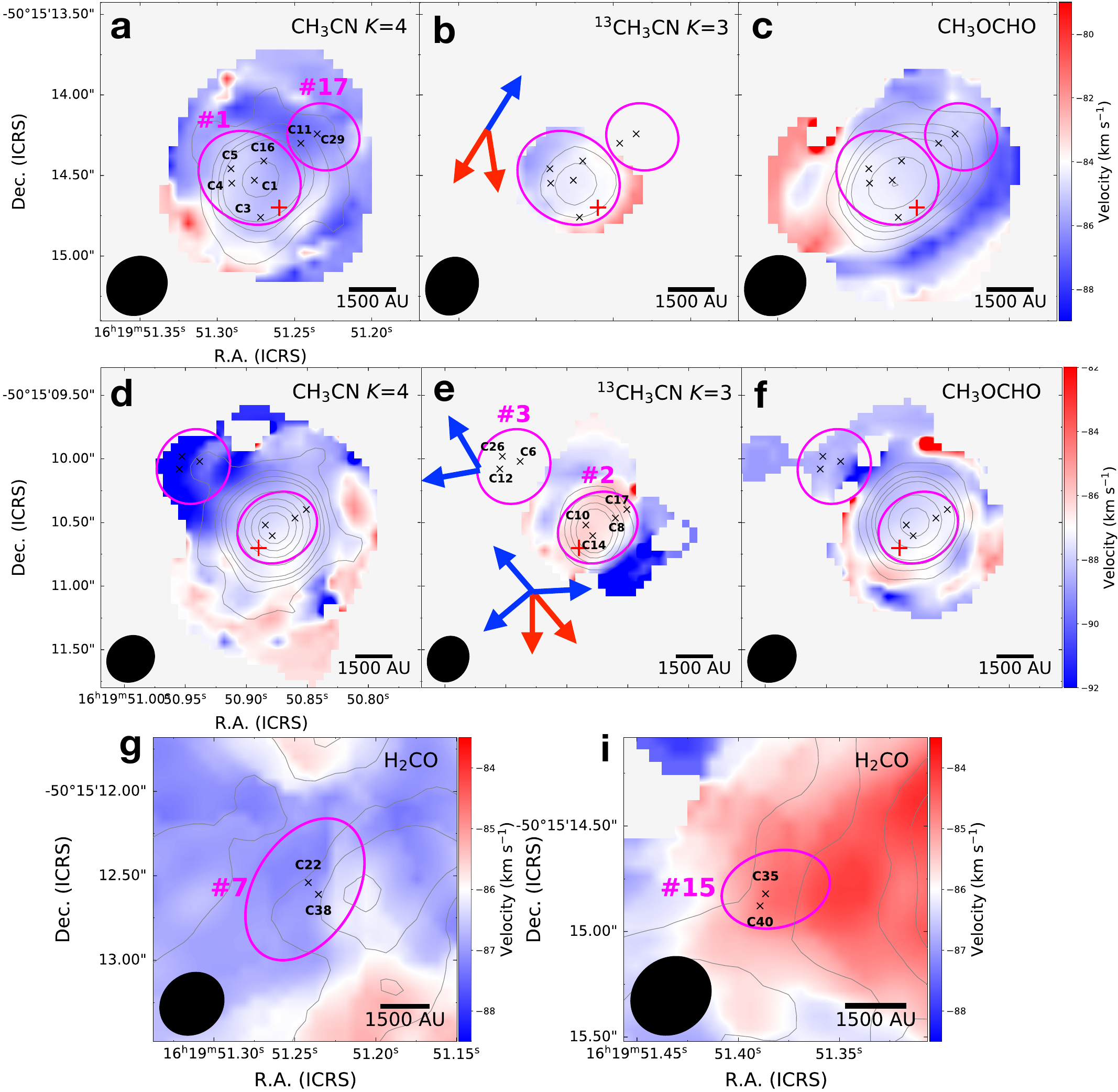}
\caption{
\textbf{Intensity-weighted velocity maps derived from the low-resolution data toward parent cores of multiple systems.} 
\textbf{a--f}, We show intensity-weighted velocity maps of CH$_3$CN $12_4-11_4$ (\textbf{a} and \textbf{d}), $^{13}$CH$_3$CN $13_3-12_3$ (\textbf{b} and \textbf{e}), and CH$_3$OCHO $20_{0,20}-19_{0,19}$ (\textbf{c} and \textbf{f}) for dense cores \#1, \#17, \#2, and \#3. 
The blue and red arrows show the directions of the outflows seen in the SiO emission from the ALMA low-resolution data.   
\textbf{g--i}, Intensity-weighted velocity maps of H$_2$CO $3_{2,2}-2_{2,1}$ for dense cores \#7 (\textbf{g}) and \#15 (\textbf{i}). 
The magenta ellipses and black crosses show the dense cores and their embedded condensations, respectively. 
The red plusses marks the Class II CH$_{3}$OH maser positions. 
The grey contours show the velocity-integrated intensity maps with levels of [5, 10, 15, 20, 40, 60, 80, 100]$\times \sigma_{\rm rms}$, where the $\sigma_{\rm rms}$ is
13~mJy~beam$^{-1}$ \kms{} for \textbf{a--f} and 9~mJy~beam$^{-1}$ \kms{} for \textbf{g--i}.  
The black ellipses in the lower left corner of each panel denote the synthesized beam of lines images.
}
\label{fig:lowmom1}
\end{figure}

\begin{figure}[!h]
\centering
\includegraphics[width=1\textwidth]{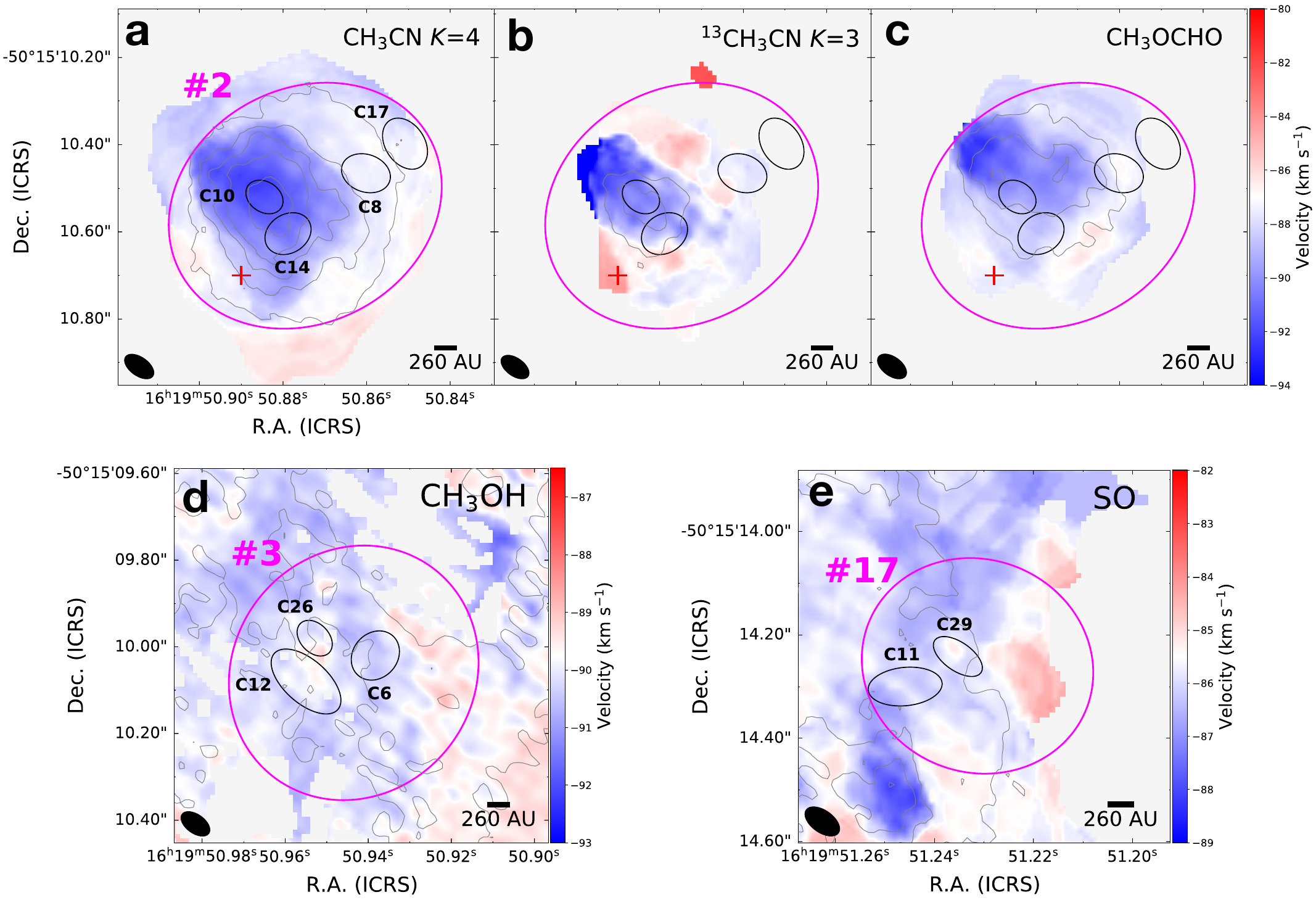}
\caption{ \textbf{Intensity-weighted velocity maps derived from the high-resolution data toward  multiple systems.} 
\textbf{a--c}, Intensity-weighted velocity maps of CH$_3$CN $12_4-11_4$ (\textbf{a}), $^{13}$CH$_3$CN $13_3-12_3$ (\textbf{b}), and CH$_3$OCHO $20_{0,20}-19_{0,19}$ (\textbf{c}) for the quadruple system. 
\textbf{d-e}, Intensity-weighted velocity maps of CH$_3$OH $4_{2,2}-3_{1,2}$ (\textbf{d}) and SO $5,6-4,5$  (\textbf{e}) for the triple and the binary systems, respectively.  
The black and magenta ellipses show the condensations and their parent cores, respectively. 
The red plusses marks the Class II  CH$_{3}$OH maser positions. 
The grey contours show the velocity-integrated intensity maps with levels of [5, 10, 15, 20, 30]$\times \sigma_{\rm rms}$, where the $\sigma_{\rm rms}$ is  9~mJy~beam$^{-1}$ \kms{} for \textbf{a--c} and 6 ~mJy~beam$^{-1}$ \kms{} for \textbf{d--e}. 
The black ellipses in the lower left corner of each panel denote the synthesized beam of lines images.
}
\label{fig:highmom1}
\end{figure}

\begin{figure}[!h]
\centering
\includegraphics[width=1\textwidth]{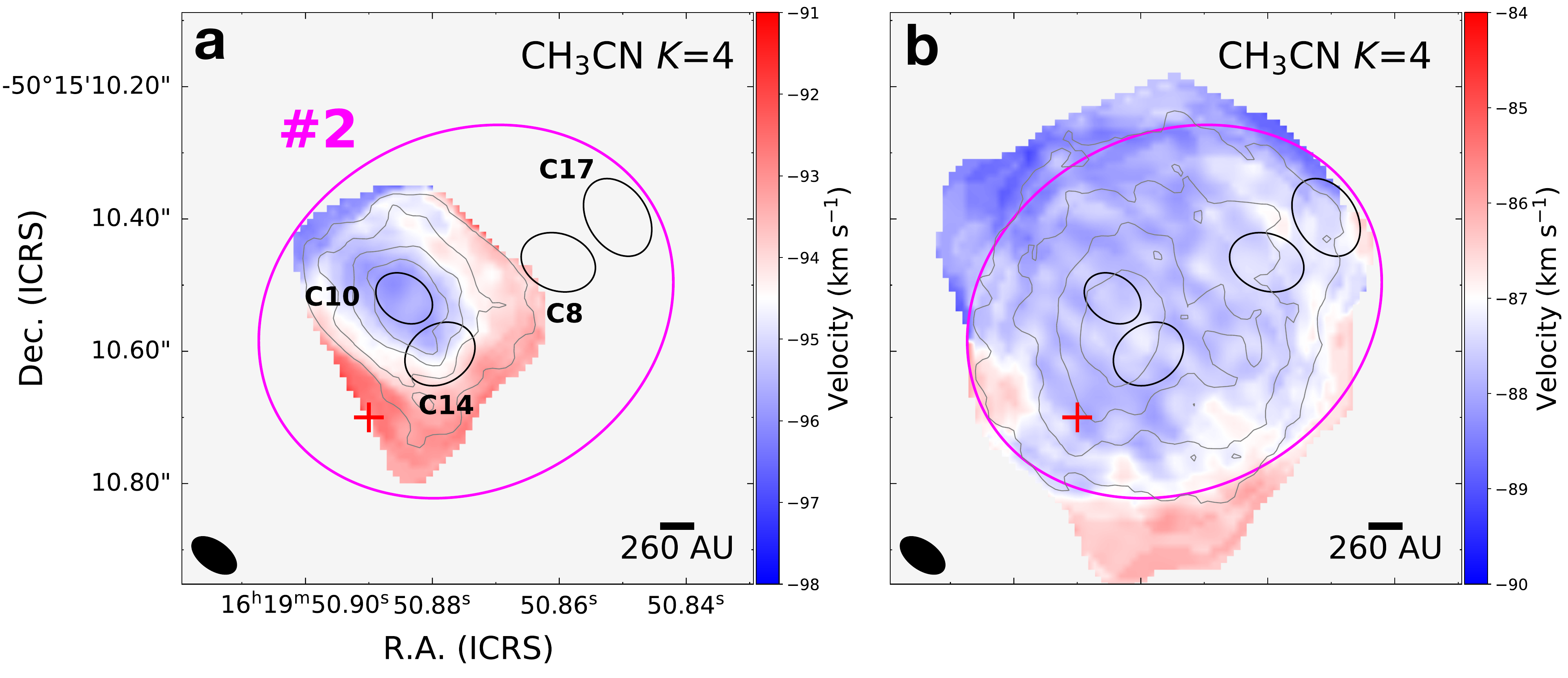}
\caption{
\textbf{Intensity-weighted velocity maps of CH$_3$CN $12_4-11_4$ for two velocity ranges of  [-98, -91] \kms\ and [-90, -84] \kms.} 
The black and magenta ellipses show the condensations and their parent cores, respectively. 
The red plusses marks the Class II  CH$_{3}$OH maser positions. 
The grey contours show the velocity-integrated intensity maps with levels of [5, 10, 15, 20, 30]$\times \sigma_{\rm rms}$, where the $\sigma_{\rm rms}$ is   6~mJy~beam$^{-1}$ \kms. 
The black ellipses in the lower left corner of each panel denote the synthesized beam of lines images. 
}
\label{fig:highmom11}
\end{figure}

\begin{figure}[!h]
\centering
\includegraphics[width=1\textwidth]{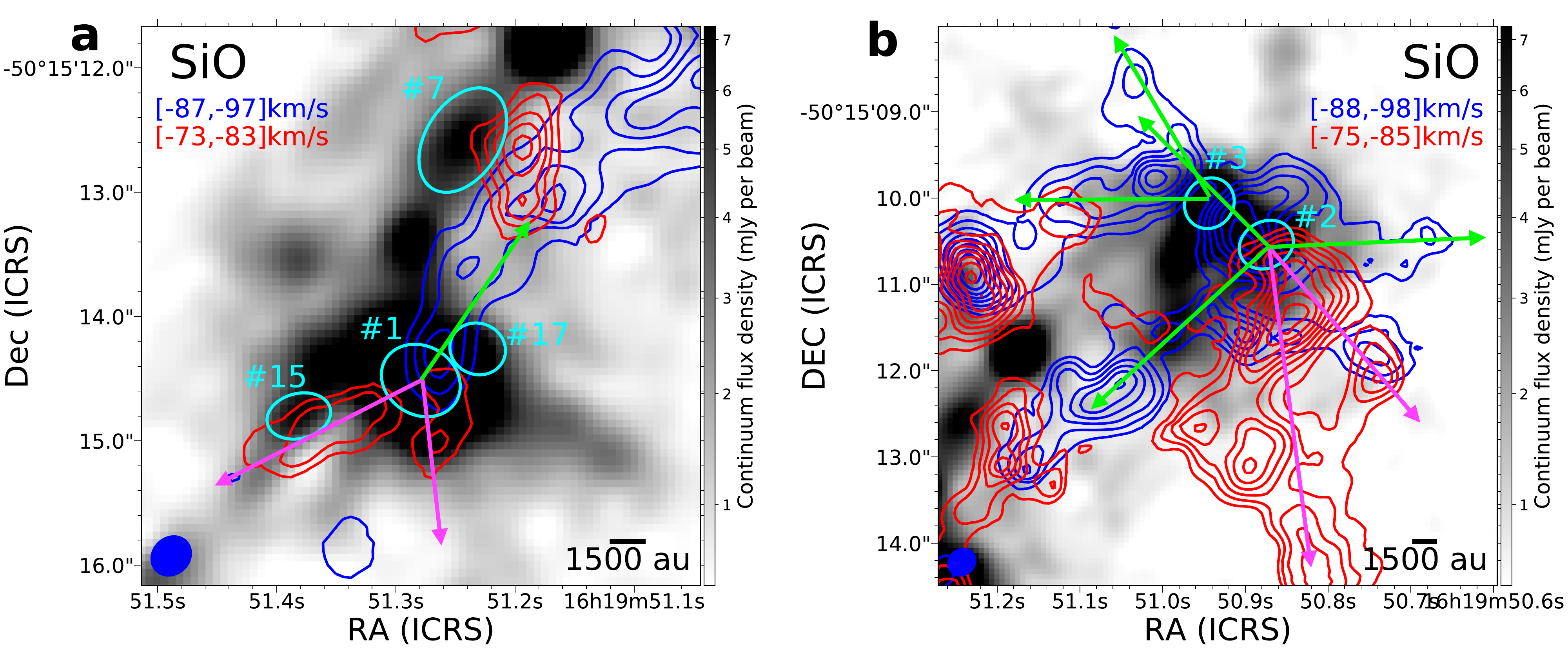}
\caption{
\textbf{Molecular outflows seen in SiO emission in the ALMA low-resolution data.} 
SiO redshifted and blueshifted contours overlaid on the ALMA low-resolution 1.3~mm continuum. The green and magenta arrows present the blueshifted and redshifted directions of the SiO outflow.  
The cyan ellipses show the dense cores.    
In panel \textbf{a}, the SiO emission shows a prominent blueshifted, a clear redshifted, and a weak redshifted component from dense core \#1. 
In panel \textbf{b}, the SiO emission reveals complicated outflow spatial morphologies toward dense cores \#2 and \#3. 
Dense core \#3 drives two blueshifted outflows. 
Dense core \#2  drives at least two blueshifted and two redshifted outflows. 
Contours levels start  at  3 $\sigma_{\rm rms}$ and increase in step of 1.5 $\sigma_{\rm rms}$ interval, where $\sigma_{\rm rms}$ is 0.023 Jy beam$^{-1}$ km s$^{-1}$. 
The synthesized beam size of 1.3 mm continuum image present in the lower left corner.  
The detected misaligned outflows further suggest that the quintuple, quadruple, and triple systems are formed from core fragmentation \citep{Offner2016}. 
}
\label{fig:outflow}
\end{figure}

\clearpage

\setcounter{figure}{0}
\renewcommand{\figurename}{Supplementary Fig.}
\renewcommand{\figureautorefname}{Supplementary Fig.}

\begin{figure}[!h]
\centering
\includegraphics[width=1\textwidth]{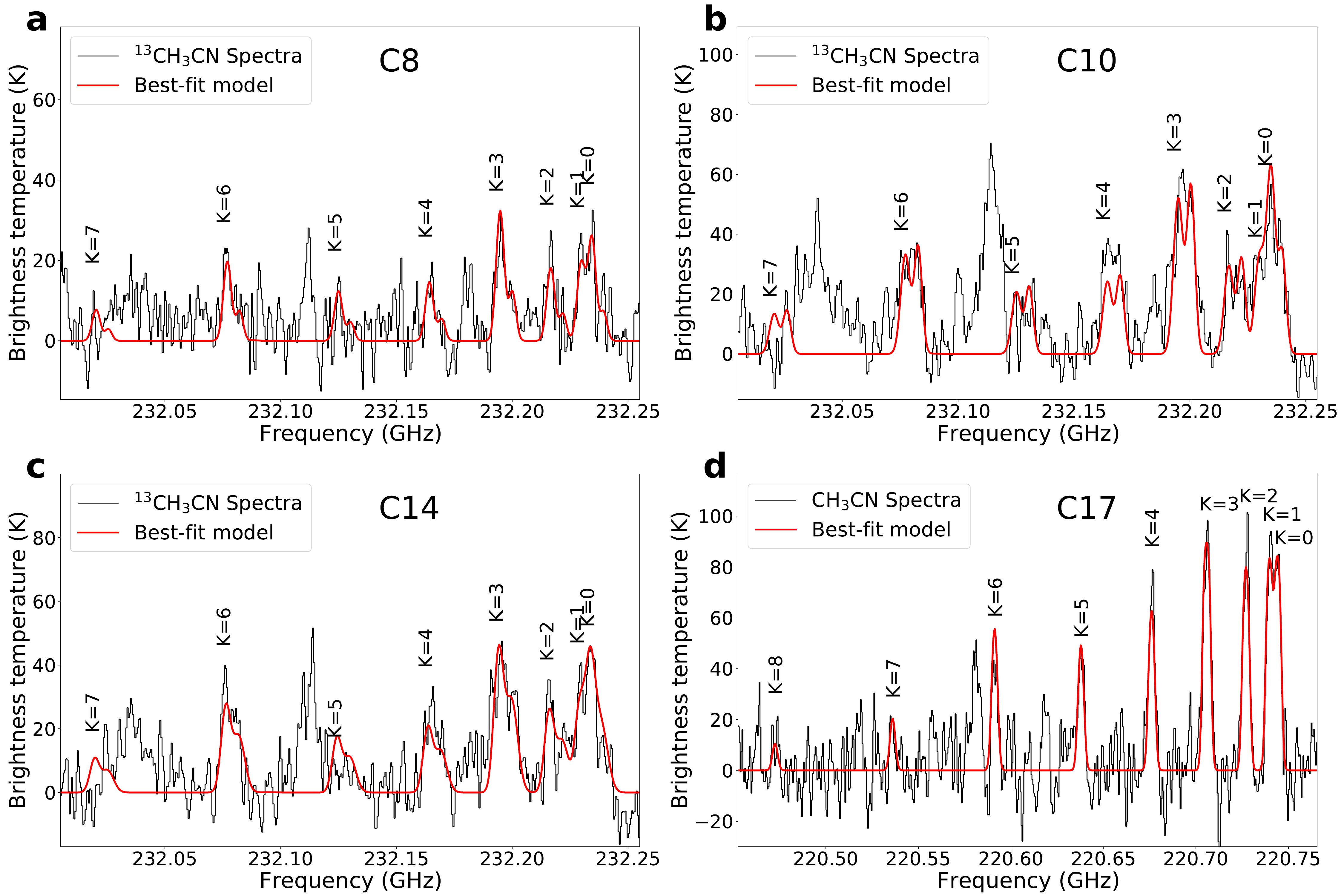}
\caption{
\textbf{Examples of spectra of  \tmthc{} and \mthc{} for the quadruple system.} 
Panels \textbf{a, b, c} show the spectra of \tmthc{} toward condensations of C8, C10 and C14, respectively.  
Panel \textbf{d} shows the spectrum of \mthc{} toward condensation C17 where the \tmthc{} is not sufficiently strong to allow a temperature measurement. 
The results of XCLASS best fit are overlaid on the spectra as red curves in each panel.  
}
\label{fig:spectra}
\end{figure}


\clearpage


\end{document}